\documentclass[twocolumn]{aastex62}

\usepackage{color}
\usepackage{amsmath}
\usepackage{graphicx}
\usepackage{natbib}
\usepackage{multirow}
\usepackage{booktabs}
\usepackage{tabularx}
\usepackage{savesym}
\savesymbol{tablenum}
\usepackage[scientific-notation=true]{siunitx} % auto scientific notation with \num{}
\restoresymbol{SIX}{tablenum}

\newcommand{\Qr}{$\mathcal{Q}_r$}
\newcommand{\Ur}{$\mathcal{U}_r$}
\newcommand{\msi}{$m_{\mathrm{Si}}$}
\newcommand{\mac}{$m_{\mathrm{aC}}$}
\newcommand{\mice}{$m_{\mathrm{H2O}}$}

\definecolor{lightgray}{gray}{0.9}

\begin{document}

%% If you wish, you may supply running head information, although
%% this information may be modified by the editorial offices.
%% The left head contains a list of authors,
%% usually a maximum of three (otherwise use et al.).  The right
%% head is a modified title of up to roughly 44 characters.
%% Running heads will not print in the manuscript style.

\shorttitle{HD 35841 Dust Ring Resolved with GPI \& STIS}
\shortauthors{Esposito et al.}


% TITLE

\title{Direct Imaging of the HD 35841 Debris Disk:\\A Polarized Dust Ring from Gemini Planet Imager and an Outer Halo from \emph{HST}/STIS}

%% Use \author, \affiliation, and the \and command to format

% AUTHORS

\correspondingauthor{Thomas M. Esposito}
\email{tesposito@berkeley.edu}

% Only format to set author list as of aastex6.1.

\author[0000-0002-0792-3719]{Thomas M. Esposito}
\affiliation{Astronomy Department, University of California, Berkeley, CA 94720, USA}

\author[0000-0002-5092-6464]{Gaspard Duch\^ene}
\affiliation{Astronomy Department, University of California, Berkeley, CA 94720, USA}
\affiliation{Universit\'e Grenoble Alpes / CNRS, Institut de Plan\'etologie et d'Astrophysique de Grenoble, 38000 Grenoble, France}

\author{Paul Kalas}
\affiliation{Astronomy Department, University of California, Berkeley, CA 94720, USA}
\affiliation{SETI Institute, Carl Sagan Center, 189 Bernardo Ave.,  Mountain View CA 94043, USA}

\author{Malena Rice}
\affiliation{Astronomy Department, University of California, Berkeley, CA 94720, USA}
\affiliation{Department of Astronomy, Yale University, New Haven, CT 06511, USA}

\author[0000-0002-9173-0740]{\'{E}lodie Choquet}
\altaffiliation{NASA Hubble Fellow}
\affiliation{Department of Astronomy, California Institute of Technology, 1200 E. California Blvd, Pasadena, CA 91125, USA}
\affiliation{NASA Jet Propulsion Laboratory, California Institute of Technology, Pasadena, CA 91109, USA}

\author[0000-0003-1698-9696]{Bin Ren}
\affiliation{Department of Physics and Astronomy, Johns Hopkins University, Baltimore, MD 21218, USA}
\affiliation{Space Telescope Science Institute, Baltimore, MD 21218, USA}

\author[0000-0002-3191-8151]{Marshall D. Perrin}
\affiliation{Space Telescope Science Institute, Baltimore, MD 21218, USA}

\author{Christine H. Chen}
\affiliation{Space Telescope Science Institute, Baltimore, MD 21218, USA}

\author[0000-0001-6364-2834]{Pauline Arriaga}
\affiliation{Department of Physics \& Astronomy, 430 Portola Plaza, University of California, Los Angeles, CA 90095, USA}

\author{Eugene Chiang}
\affiliation{Astronomy Department, University of California, Berkeley, CA 94720, USA}
\affiliation{Earth and Planetary Science Department, University of California, Berkeley, CA 94720, USA}

\author[0000-0001-6975-9056]{Eric L. Nielsen}
\affiliation{SETI Institute, Carl Sagan Center, 189 Bernardo Ave.,  Mountain View CA 94043, USA}
\affiliation{Kavli Institute for Particle Astrophysics and Cosmology, Stanford University, Stanford, CA 94305, USA}

\author{James R. Graham}
\affiliation{Astronomy Department, University of California, Berkeley, CA 94720, USA}

\author[0000-0003-0774-6502]{Jason J. Wang}
\affiliation{Astronomy Department, University of California, Berkeley, CA 94720, USA}

\author[0000-0002-4918-0247]{Robert J. De Rosa}
\affiliation{Astronomy Department, University of California, Berkeley, CA 94720, USA}

\author[0000-0002-7821-0695]{Katherine B. Follette}
\altaffiliation{NASA Sagan Fellow}
\affiliation{Kavli Institute for Particle Astrophysics and Cosmology, Stanford University, Stanford, CA 94305, USA}
\affiliation{Physics and Astronomy Department, Amherst College, 21 Merrill Science Drive, Amherst, MA 01002, USA}

\author[0000-0001-5172-7902]{S. Mark Ammons}
\affiliation{Lawrence Livermore National Laboratory, 7000 East Ave, Livermore, CA 94550, USA}

\author[0000-0003-4142-9842]{Megan Ansdell}
\affiliation{Astronomy Department, University of California, Berkeley, CA 94720, USA}

\author[0000-0002-5407-2806]{Vanessa P. Bailey}
\affiliation{NASA Jet Propulsion Laboratory, California Institute of Technology, Pasadena, CA 91109, USA}

\author[0000-0002-7129-3002]{Travis Barman}
\affiliation{Lunar and Planetary Laboratory, University of Arizona, Tucson AZ 85721, USA}

\author{Juan Sebasti{\'a}n Bruzzone}
\affiliation{Department of Physics and Astronomy, The University of Western Ontario, London, ON, N6A 3K7, Canada}

\author{Joanna Bulger}
\affiliation{Subaru Telescope, NAOJ, 650 North A{'o}hoku Place, Hilo, HI 96720, USA}

\author[0000-0001-6305-7272]{Jeffrey Chilcote}
\affiliation{Kavli Institute for Particle Astrophysics and Cosmology, Stanford University, Stanford, CA 94305, USA}
\affiliation{Department of Physics, University of Notre Dame, 225 Nieuwland Science Hall, Notre Dame, IN, 46556, USA}

\author[0000-0003-0156-3019]{Tara Cotten}
\affiliation{Department of Physics and Astronomy, University of Georgia, Athens, GA 30602, USA}

\author{Rene Doyon}
\affiliation{Institut de Recherche sur les Exoplan{\`e}tes, D{\'e}partement de Physique, Universit{\'e} de Montr{\'e}al, Montr{\'e}al QC, H3C 3J7, Canada}

\author[0000-0002-0176-8973]{Michael P. Fitzgerald}
\affiliation{Department of Physics \& Astronomy, 430 Portola Plaza, University of California, Los Angeles, CA 90095, USA}

\author[0000-0002-4144-5116]{Stephen J. Goodsell}
\affiliation{Gemini Observatory, 670 N. A'ohoku Place, Hilo, HI 96720, USA}

\author[0000-0002-7162-8036]{Alexandra Z. Greenbaum}
\affiliation{Department of Astronomy, University of Michigan, Ann Arbor, MI 48109, USA}

\author[0000-0003-3726-5494]{Pascale Hibon}
\affiliation{Gemini Observatory, Casilla 603, La Serena, Chile}

\author[0000-0003-1498-6088]{Li-Wei Hung}
\affiliation{Department of Physics \& Astronomy, 430 Portola Plaza, University of California, Los Angeles, CA 90095, USA}

\author{Patrick Ingraham}
\affiliation{Large Synoptic Survey Telescope, 950N Cherry Ave., Tucson, AZ 85719, USA}

\author[0000-0002-9936-6285]{Quinn Konopacky}
\affiliation{Center for Astrophysics and Space Science, University of California San Diego, La Jolla, CA 92093, USA}

\author{James E. Larkin}
\affiliation{Department of Physics \& Astronomy, 430 Portola Plaza, University of California, Los Angeles, CA 90095, USA}

\author[0000-0003-1212-7538]{Bruce Macintosh}
\affiliation{Kavli Institute for Particle Astrophysics and Cosmology, Stanford University, Stanford, CA 94305, USA}

\author{J\'er\^ome Maire}
\affiliation{Center for Astrophysics and Space Science, University of California San Diego, La Jolla, CA 92093, USA}

\author[0000-0001-7016-7277]{Franck Marchis}
\affiliation{SETI Institute, Carl Sagan Center, 189 Bernardo Ave.,  Mountain View CA 94043, USA}

\author[0000-0002-4164-4182]{Christian Marois}
\affiliation{National Research Council of Canada Herzberg, 5071 West Saanich Rd, Victoria, BC, V9E 2E7, Canada}
\affiliation{University of Victoria, 3800 Finnerty Rd, Victoria, BC, V8P 5C2, Canada}

\author[0000-0002-9133-3091]{Johan Mazoyer}
\affiliation{Department of Physics and Astronomy, Johns Hopkins University, Baltimore, MD 21218, USA}
\affiliation{Space Telescope Science Institute, Baltimore, MD 21218, USA}

\author[0000-0003-3050-8203]{Stanimir Metchev}
\affiliation{Department of Physics and Astronomy, Centre for Planetary Science and Exploration, The University of Western Ontario, London, ON N6A 3K7, Canada}
\affiliation{Department of Physics and Astronomy, Stony Brook University, Stony Brook, NY 11794-3800, USA}

\author[0000-0001-6205-9233]{Maxwell A. Millar-Blanchaer}
\altaffiliation{NASA Hubble Fellow}
\affiliation{NASA Jet Propulsion Laboratory, California Institute of Technology, Pasadena, CA 91109, USA}

\author[0000-0001-7130-7681]{Rebecca Oppenheimer}
\affiliation{Department of Astrophysics, American Museum of Natural History, New York, NY 10024, USA}

\author{David Palmer}
\affiliation{Lawrence Livermore National Laboratory, 7000 East Ave, Livermore, CA 94550, USA}

\author{Jennifer Patience}
\affiliation{School of Earth and Space Exploration, Arizona State University, PO Box 871404, Tempe, AZ 85287, USA}

\author{Lisa Poyneer}
\affiliation{Lawrence Livermore National Laboratory, 7000 East Ave, Livermore, CA 94550, USA}

\author{Laurent Pueyo}
\affiliation{Space Telescope Science Institute, Baltimore, MD 21218, USA}

\author[0000-0002-9246-5467]{Abhijith Rajan}
\affiliation{School of Earth and Space Exploration, Arizona State University, PO Box 871404, Tempe, AZ 85287, USA}

\author[0000-0003-0029-0258]{Julien Rameau}
\affiliation{Institut de Recherche sur les Exoplan{\`e}tes, D{\'e}partement de Physique, Universit{\'e} de Montr{\'e}al, Montr{\'e}al QC, H3C 3J7, Canada}

\author[0000-0002-9667-2244]{Fredrik T. Rantakyr\"o}
\affiliation{Gemini Observatory, Casilla 603, La Serena, Chile}

\author{Dominic Ryan}
\affiliation{Astronomy Department, University of California, Berkeley, CA 94720, USA}

\author[0000-0002-8711-7206]{Dmitry Savransky}
\affiliation{Sibley School of Mechanical and Aerospace Engineering, Cornell University, Ithaca, NY 14853, USA}

\author{Adam C. Schneider}
\affiliation{School of Earth and Space Exploration, Arizona State University, PO Box 871404, Tempe, AZ 85287, USA}

\author[0000-0003-1251-4124]{Anand Sivaramakrishnan}
\affiliation{Space Telescope Science Institute, Baltimore, MD 21218, USA}

\author[0000-0002-5815-7372]{Inseok Song}
\affiliation{Department of Physics and Astronomy, University of Georgia, Athens, GA 30602, USA}

\author[0000-0003-2753-2819]{R\'{e}mi Soummer}
\affiliation{Space Telescope Science Institute, Baltimore, MD 21218, USA}

\author[0000-0002-9121-3436]{Sandrine Thomas}
\affiliation{Large Synoptic Survey Telescope, 950N Cherry Ave., Tucson, AZ 85719, USA}

\author{J. Kent Wallace}
\affiliation{NASA Jet Propulsion Laboratory, California Institute of Technology, Pasadena, CA 91109, USA}

\author[0000-0002-4479-8291]{Kimberly Ward-Duong}
\affiliation{School of Earth and Space Exploration, Arizona State University, PO Box 871404, Tempe, AZ 85287, USA}
\affiliation{Physics and Astronomy Department, Amherst College, 21 Merrill Science Drive, Amherst, MA 01002, USA}

\author[0000-0003-4483-5037]{Sloane Wiktorowicz}
\affiliation{Department of Astronomy, UC Santa Cruz, 1156 High St., Santa Cruz, CA 95064, USA }

\author[0000-0002-9977-8255]{Schuyler Wolff}
\affiliation{Leiden Observatory, Leiden University, P.O. Box 9513, 2300 RA Leiden, The Netherlands}


% ABSTRACT
\begin{abstract}

We present new high resolution imaging of a light-scattering dust ring and halo around the young star HD 35841. Using spectroscopic and polarimetric data from the Gemini Planet Imager in $H$-band (1.6 $\micron$), we detect the highly inclined ($i=85\degr$) ring of debris down to a projected separation of ${\sim}12$ au (${\sim}0\farcs12$) for the first time. Optical imaging from \emph{HST}/STIS shows a smooth dust halo extending outward from the ring to $>$140 au (${>}1.4\arcsec$). We measure the ring's scattering phase function and polarization fraction over scattering angles of $22\degr$--$125\degr$, showing a preference for forward scattering and a polarization fraction that peaks at ${\sim}30$\% near the ansae. Modeling of the scattered-light disk indicates that the ring spans radii of ${\sim}60$--220 au, has a vertical thickness similar to that of other resolved dust rings, and contains grains as small as 1.5 $\micron$ in diameter. These models also suggest the grains have a low porosity, are more likely to consist of carbon than astrosilicates, and contain significant water ice. The halo has a surface brightness profile consistent with that expected from grains pushed by radiation pressure from the main ring onto highly eccentric but still bound orbits. We also briefly investigate arrangements of a possible inner disk component implied by our spectral energy distribution models, and speculate about the limitations of Mie theory for doing detailed analyses of debris disk dust populations.

\end{abstract}

% KEYWORDS
%% Keywords should appear after the \end{abstract} command. The uncommented example has been keyed in ApJ style. See the instructions to authors for the journal to which you are submitting your paper to determine what keyword punctuation is appropriate.

\keywords{circumstellar matter - infrared: planetary systems - stars: individual (HD 35841) - techniques: high angular resolution}

% INTRODUCTION
\section{INTRODUCTION} \label{sect:intro}

Recent advances in high-contrast imaging have offered direct observations of inner planetary systems that were previously inaccessible. In particular, we can now resolve circumstellar debris disk components with smaller radii and lower surface brightnesses than the bright, extended components discovered in the last decade. The near-infrared signatures of these disks are produced by micron-sized grains of rock and ice that scatter light from the host star. These grains are collisional products of larger bodies in the system. Together, these materials represent the building blocks and leftovers of planet formation, giving us an indirect probe of planetary system evolution.

HD 35841 is an F5V star at a distance of $102.9\pm4.2$ pc \citep{astraatmadja2016, gaia_mission} and is a purported member of the Columba moving group \citep{moor2006, torres2008}, giving it an age of $\sim$40 Myr \citep{bell2015}. The star's infrared excess was first noted by \citet{silverstone2000} with $L_{IR}/L_*\approx\num{1.3e-3}$. A corresponding dust disk was later resolved in archival \emph{Hubble Space Telescope} (HST) NICMOS 1.1-$\micron$ data by \citet{soummer2014}. The nearly edge-on disk was detected out to $1\farcs5$ (${\sim}150$ au with our updated distance) in radius and showed an apparent wing-tilt asymmetry where the position angles of the midplanes of the two sides of the disk are $\sim$25$\degr$ from being diametrically opposed, a much greater tilt than the few degrees observed in $\beta$ Pic's \citep{kalas1995}. However, image resolution was limited to ${\sim}0\farcs1$ and no information was available interior to ${\sim}0\farcs3$.

We present new Gemini Planet Imager (GPI; \citealt{macintosh2014}) $H$-band data that resolve the disk into a well-defined ring for the first time and provide the first polarized intensity image. We detect the ring at a diffraction-limited resolution of ${\sim}0\farcs04$ down to a separation of $0\farcs12$ (12 au). From these images we extract scattering phase functions and polarization fractions. We also present new \emph{HST} STIS data that show the outer disk at optical wavelengths with spatial resolution of ${\sim}0\farcs05$ at separations $>0\farcs5$. Combining the GPI and STIS data, we compute an optical vs. near-IR color for the disk. Using the data from both instruments for comparison, we construct disk models that partially constrain the composition and location of the dust responsible for the scattered-light profiles. Addtionally, we compare the resulting model spectral energy distribution (SED) to existing photometry to investigate the possibility of multiple dust populations contributing to disk flux at different wavelengths.

% ROADMAP
In the following sections, we provide details about our observations and data reduction methods (Section \ref{sect:obs}), and present measurements of disk properties from our images (Section \ref{sect:img_results}). Then we describe modeling of the disk to infer its physical parameters (Section \ref{sect:modeling}). Finally, we discuss the implications of our results in broader context (Section \ref{sect:discussion}) and summarize our conclusions (Section \ref{sect:conclusions}).

\pagebreak % for latex arrangement
% OBSERVATIONS AND REDUCTION
\section{OBSERVATIONS AND DATA REDUCTION} \label{sect:obs}

% GPI OBSERVATIONS AND REDUCTION
\subsection{Gemini Planet Imager} \label{sect:gpi_obs}

We observed HD 35841 with GPI in {\it H}-band ($\lambda_{cen}=1.647$ \micron) using its spectroscopic (``H-Spec'') and polarimetric (``H-Pol'') modes as part of the Gemini Planet Imager Exoplanet Survey (GPIES; PI B. Macintosh). Details of the data sets are listed in Table \ref{tab:obs_specs}. In both cases, the pixel scale was 14.166 $\pm$ 0.007 mas $\mathrm{lenslet^{-1}}$ \citep{derosa2015b}, a 123 mas radius focal plane mask (FPM) occulted the star, and angular differential imaging (ADI; \citealt{marois2006}) was employed (the default for GPI). The average atmospheric properties for the H-Spec data set were: DIMM seeing = $1\farcs0\pm0\farcs2$, MASS seeing = $0\farcs5\pm0\farcs1$, coherence time $\tau=5.4\pm1.2$ ms, and airmass ranged from 1.01 to 1.06. For H-Pol, the airmass ranged from 1.08 to 1.19 but atmospheric measurements were not available from the observatory.

% Observations table.
\begin{table}[ht]
\begin{center}
\caption{HD 35841 Observations}
\label{tab:obs_specs}

\begin{tabular}{l c c c c c}
\toprule
Inst./Mode & Filter & $t_\mathrm{exp}$ & $N_\mathrm{exp}$ & $\Delta$PA & Date  \\
&  & (s) &  & (deg) &   \\
\midrule
GPI/Spec & $H$ & 59.65 & 44 & 32.1 & 2016/2/28 \\
GPI/Pol & $H$ & 88.74 & 28 & 3.8 & 2016/3/18 \\
STIS/A0.6 & 50CCD & 120.0 & 12 & 16* & 2014/11/6 \\
STIS/A1.0 & 50CCD & 485.0 & 6 & 16* & 2014/11/6 \\
\bottomrule

\end{tabular}
\end{center}

$t_{exp}$ is the exposure time per image and $N_{exp}$ is the total number of images in a given mode.\\
* The STIS $\Delta$PA is comprised of only two roll angles.

\end{table}

The spectroscopic observations divide the filter bandpass into micro-spectra that are measured by the detector and then converted into 37 spectral channels per image by the GPI Data Reduction Pipeline (DRP; \citealt{perrin2014_drp, perrin2016_drp}). We used this pipeline's standard methods to assemble the raw data into 44 spectral data cubes (similar to steps taken in \citealt{derosa2016}). The star location in each channel was determined from measurements of the four fiducial ``satellite'' spots \citep{sivaramakrishnan2006, wang2014, pueyo2015}, as was the photometric calibration, assuming a satellite-spot-to-star flux ratio of \num{2.035e-4} \citep{maire2014} and a stellar $H$ magnitude of $7.842\pm0.034$ (Two Micron All Sky Survey [2MASS], \citealt{cutri2003}). We did not high-pass filter or smooth any of the GPI images used for our measurements and analysis. In this paper we consider only the broadband-collapsed results from the spectral data; any consideration of the disk's spectrum in reflected light is left for a future work.

We applied multiple techniques to subtract the stellar point-spread function (PSF). First, we used \texttt{pyKLIP}, a Python implementation \citep{wang2015_pyklip} of the Karhunen--Lo\`eve Image Projection (KLIP) algorithm \citep{soummer2012, pueyo2016}. Subtraction was performed on a channel-by-channel basis using only angular diversity of reference images (no spectral diversity was used) and the aggressiveness of the PSF subtraction was adjusted by varying the KLIP parameters. We show aggressive and conservative reductions in Figure \ref{fig:imstack}. The aggressive reduction divided each image radially into 8 equal-width annuli between $r=10$ and 85 pixels (px) with no azimuthal division of the annuli, used a minimum rotation threshold of 1 px to select reference images, and projected onto the first 44 KL modes. The conservative reduction was identical except it employed only the first KL mode. The PSF-subtracted images were derotated so north is up and averaged into the final image.

When using the aggressive pyKLIP parameters with the entire 44-image data set, we found the PSF subtraction to preferentially self-subtract the ring along its most southeast edge \citep{milli2012}. No such effect was found for the conservative reduction. The effect possibly arose because more images were taken after transit than before transit, leading to an unequal distribution of the disk's position angle (PA) among reference images; we could avoid the bias with the aggressive parameters by using a subset of 30 images that was balanced in number of images pre- and post-transit. However, this resulted in lower S/N for the rest of the ring, so we choose to present the 44-image version in Figure \ref{fig:imstack} to better display the ring's other features and illustrate this phenomenon.

Separately, we used a modified version of the ``locally optimized combination of images'' (LOCI) algorithm \citep{lafreniere2007} on images that were median-collapsed across spectral channels. This collapse allowed faster forward-modeling of the disk self-subtraction \citep{esposito2014} during the modeling discussed in Section \ref{sect:modeling} but did reduce S/N compared to non-collapsed reductions. The reduction presented herein used only one subtraction annulus at $r=9$--$120$ px divided azimuthally into three subsections, with LOCI parameter values of $N_{\delta}=0.5$, $W=4$\,px, $dr=120$\,px, $g=0.625$, and $N_a=160$, following the conventional definitions in \citet{lafreniere2007}. To prevent speckle noise at the edge of the FPM from detrimentally biasing subtraction over the entire annulus, we set the inner radius of the region used to optimize the LOCI coefficients to 12 px instead of 9 px. We found the preferential self-subtraction of the southeast edge noted above to also occur here with a 44-image data set, so we used the more PA-balanced subset of 30 images instead. Finally, the PSF-subtracted frames were rotated to place north up and collapsed into the final median image shown in Figure \ref{fig:imstack}.

The polarimetric data were created with GPI's Wollaston prism, which splits the light from the spectrograph's lenslets into two orthogonal polarization states. To combine the raw data into a set of polarization datacubes, we used the GPI DRP with the methods outlined in \citet{esposito2016} and described in more detail in \citet{perrin2015_4796} and \citet{millar-blanchaer2015}. Specific to this data set, the instrumental polarization was removed by first estimating the apparent stellar polarization in each datacube as the mean normalized difference of pixels 2--11 px from the star's location (i.e., both inside and just outside of the FPM). For each pixel in the cube, we then scaled that value by the pixel's total intensity and subtracted the product. The datacubes were also corrected for geometric distortion and smoothed with a Gaussian kernal ($\sigma=2$ px). Combining the datacubes results in a three-dimensional Stokes cube containing the Stokes parameters \{$I$, $Q$, $U$, $V$\}, rotated to place North up. Finally, the Stokes cube was photometrically calibrated using the satellite spot fluxes by assuming the same satellite-spot-to-star flux ratio as we did for H-Spec and following the methods described in \citet{hung2015b}. We recovered only a very low S/N total intensity detection from this H-Pol data set with $3.8\degr$ of field rotation, so we use only the H-Spec data for total intensity analysis.

% STIS OBSERVATIONS AND REDUCTION
\subsection{\emph{HST}/STIS} \label{sect:stis_obs}

We observed HD~35841 on 2014 November 6 with the STIS instrument on HST in its coronagraphic mode (program GO-13381, PI M.~Perrin). The system was observed at two telescope roll orientations of -78.9$\degr$ and -94.9$\degr$ over two orbits. These orientations were chosen to align the disk's major axis, estimated by \citet{soummer2014}, perpendicular to the occulting wedge for one of the rolls, but it was ultimately offset by 24$\degr$ due to scheduling constraints. At each orientation, we acquired a series of six 120-s exposures with the star centered on the 0$\farcs$6-wide WEDGEA0.6 wedge position (hereafter A0.6), then three longer 485-s exposures on the 1$\arcsec$-wide WEDGEA1.0 position (hereafter A1.0). This resulted in a total exposure time of 1,440~s for separations of $0\farcs3$--$0\farcs5$ but 4,350~s for separations $>0\farcs5$.

The STIS coronagraphic mode is unfiltered (``50CCD'') and sensitive from ${\sim}0.20$--1.03 $\micron$ with a pivot wavelength of $\lambda_p=0.5754~\micron$ \citep{stis2017}. The pixel scale is 50.77 mas $\mathrm{pixel^{-1}}$ \citep{schneider2016}.

To subtract the stellar PSF from the science images, we also observed HD 37002 as a reference star at a single orientation during the single orbit between visits to HD 35841. This minimized potential PSF differences caused by telescope thermal breathing. HD 37002 is an F5V star chosen for its close spectral match, similar luminosity, and on-sky proximity to HD 35841. We acquired six 110-s exposures on A0.6 and three 505-s exposures on A1.0.

We processed the A0.6 and A1.0 data sets separately using classical reference star differential imaging with the following steps. After flat-fielding via the STIS \texttt{calstis} pipeline and correction of the bad pixels, we registered and scaled the images of the target and reference star by minimizing the quadratic difference between each of them and the first reference image. The star center is estimated from cross-correlation of the secondary mirror struts diffraction spikes, with the absolute star center determined from a Radon transform of the first reference image \citep{pueyo2015}. For each science frame, we subtracted either the closest reference frame or the median of all reference frames, choosing the version that minimized PSF residuals. Finally, we consolidated the PSF-subtracted frames for both wedges into one pool, rotated them to set north up, masked the areas covered by the wedges and diffraction spikes, and average-combined all of the frames using a pixel-weighted combination of their respective exposure times.

After combination, we found that some stellar background remained that was approximately azimuthally symmetric. To remove this background, we fit a sixth-order polynomial to the median radial profile measured within PA = 30$\degr$--140$\degr$ (avoiding the disk), and subtracted that polynomial function from the image at all radii. We use the resulting image for all analysis apart from one instance in Section \ref{sect:halo}.

\begin{figure*}[ht!]
\centering
\includegraphics[width=6.7in]{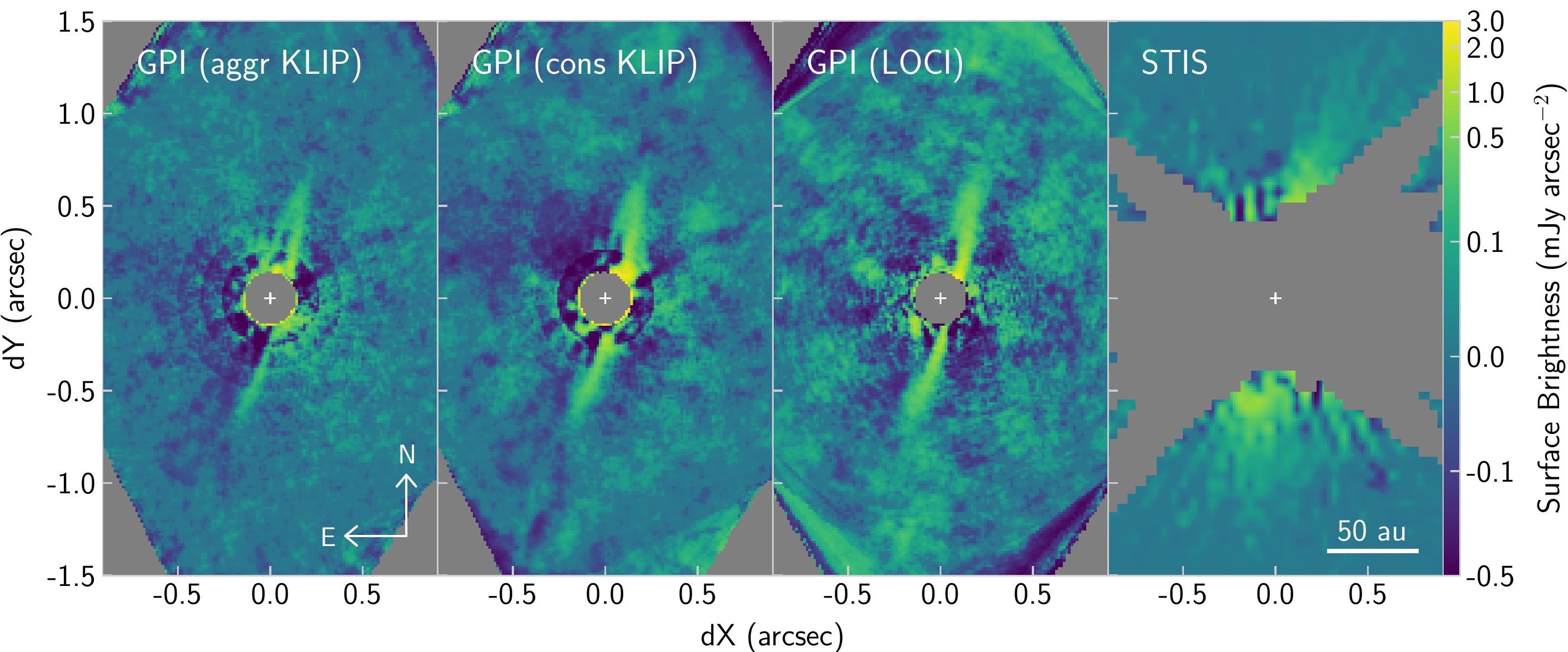}
\caption{GPI spectroscopic mode $H$-band and STIS broadband optical images on logarithmic brightness scales. The left two panels show aggressive and conservative KLIP PSF subtractions, while the third panel is the LOCI PSF subtraction. The STIS image was interpolated to match the pixel scale of the GPI images. The white cross denotes the star. Gray regions are those obscured by occulting masks, interior to our PSF-subtraction inner working angle, or outside the GPI FOV.}
\label{fig:imstack}
\end{figure*}

We converted the final image to surface brightness units of mJy arcsec$^{-2}$ via the average ``PHOTFLAM'' header value of \num{4.116e-19} erg cm$^{-2}$ s$^{-1}$ \AA$^{-1}$ and the pixel scale. For comparison with the GPI images, we linearly interpolated the final STIS image to match the GPI plate scale, and this is the version shown in Figure \ref{fig:imstack}.

% NIRC2 OBSERVATIONS AND REDUCTION
\subsection{Keck NIRC2} \label{sect:nirc2_obs}

We also reduced a Keck/NIRC2 ADI $H$-band data set from 2014 Feb 08 but failed to detect the disk with statistical significance. The data set comprised 97 frames of 20.0 s integrations with the 400-mas diameter coronagraph in place, totaling 13.9$^\circ$ of parallactic rotation. Unfortunately, data quality was suboptimal due to high humidity (${\sim}70$\%) and variable seeing of 0.6\arcsec--0.8\arcsec. The compact angular extent of this disk, combined with observing conditions and a total integration time of only 32 minutes, meant the signal was not recovered from the residual speckle noise despite attempts with classical ADI (using a median-collapsed reference PSF), LOCI, and KLIP subtractions (see Appendix \ref{sect:app_nirc2} for details).

% Total Intensity Log images.

% Polarized Intensity Log images.
\begin{figure*}[ht!]
\centering
\includegraphics[width=5.8in]{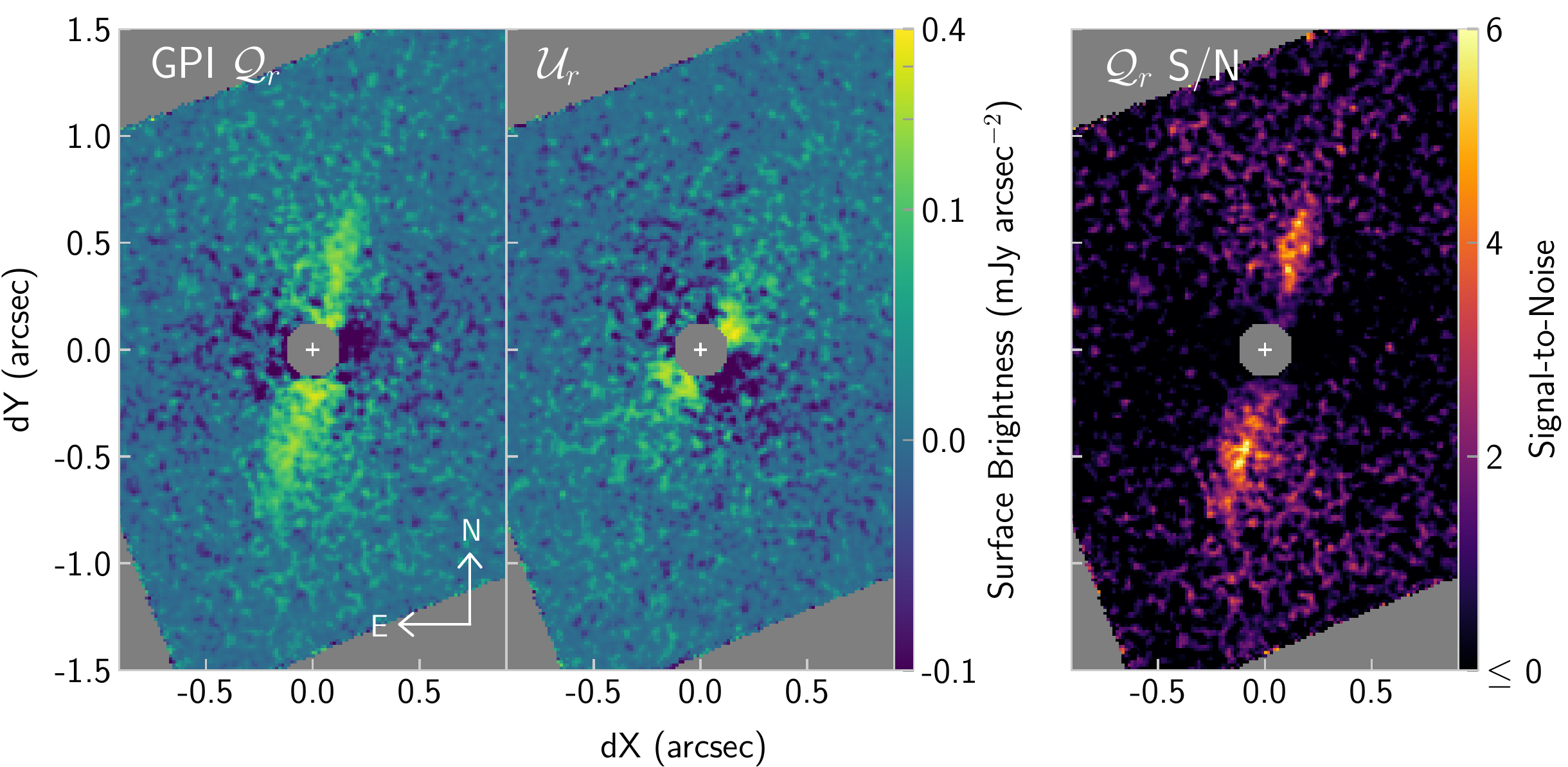}
\caption{Radial Stokes $\mathcal{Q}$ and $\mathcal{U}$ images on logarithmic brightness scales, along with the ratio of \Qr\ to a noise map derived from \Ur. The white cross denotes the star. Gray regions are those obscured by the GPI FPM or outside the FOV.}
\label{fig:polstack}
\end{figure*}

\section{OBSERVATIONAL RESULTS}

% IMAGING RESULTS
\subsection{GPI and STIS Images} \label{sect:img_results}

We present our total intensity GPI (spectroscopic mode) and STIS images in Figure \ref{fig:imstack}. The GPI images represent the aggressive KLIP, conservative KLIP, and LOCI reductions. They show a highly inclined ring of dust with sharp inner and outer edges. We assume the ring to be approximately circular, as both ansae extend to the same projected separation (${\sim}0\farcs65$ with brightness ${\sim}3\sigma$ above the local background noise) and there is no obvious stellocentric offset along the minor axis. We consider the ansae to be the portions of the ring near its intersection with the projected major axis, i.e., the inflection point of the ellipse.

The strong brightness asymmetry in the aggressive KLIP image is a reduction artifact, as the higher KL modes preferentially self-subtract the southeast edge due to the imbalance of reference image PA's previously discussed. Therefore, we use the conservative KLIP and LOCI images as the bases for our measurements and analyses. Nevertheless, we present the aggressive KLIP image because it provides the best view of the NW back edge, which is swamped by speckled residuals in conservative PSF subtractions. Additionally, bright spots along the major axis on both sides of the star at the inner working angle are likely speckle residuals, rather than point sources or ansae of an inner ring.

Photometrically, the west edge of the ring (PA 166--346$\degr$, measured east of north) appears consistently brighter than the east edge. From here on, we consider this west edge to be the ``front edge'' between the star and the observer, and the east edge to be the ``back edge'' behind the star. We base this on assumptions that the dust grains are primarily forward-scattering, their scattering properties are constant around the ring, and the ring is approximately azimuthally symmetric. While the back edge is intrinsically fainter due to a forward-scattering phase function, it is also artificially dimmed by self-subtraction by the brighter front edge. We correct for this bias in our measurements and modeling but not in the images shown in Figure \ref{fig:imstack}.

The outer edges of the ansae extend symmetrically to projected separations of $r_{proj}\approx67$ au ($0\farcs65$). We detect the ring down to our inner working angle of $r_{proj}\approx12$ au ($0\farcs12$) along the front edge, but the back edge is only marginally detected above the speckle noise level at $r_{proj}\approx27$ au ($0\farcs26$). Interior to ${\sim}19$ au ($0\farcs18$), the residual speckle noise is substantial and reduces the significance of our detection. The ring appears generally smooth, without clumps or vertical protrusions.

The STIS image is heavily impacted by the combined orientations of the occulting wedges in the constituent frames. Consequently, we detect just the ring ansae (and only partially in the NW). However, we also detect an asymmetric, low surface brightness component that extends at least $0\farcs7$ outward from the ring's outer edge and is preferentially seen west of the star. This smooth halo or ``dust apron'' becomes fainter with separation and reaches the background limit at $r_{proj}\approx140$ au ($1\farcs4$). It is the likely source of the wing-tilt asymmetry in the \citet{soummer2014} NICMOS 1.1-$\micron$ data, with forward scattering grains leading to preferential brightening of the disk's west side and creating an apparent deflection of the ansae toward that direction. This halo is reminiscent of similar features seen, for example, in the Fomalhaut, AU Mic, HD 32297, and HD 15745 disks \citep{kalas2005_fomb, chiang2009, strubbe2006, schneider2014}, but it is not sharply deflected from the main ring like the HD 61005 disk \citep{schneider2014}.

We present the polarized intensity GPI data in Figure \ref{fig:polstack} as the radial components \Qr\ and \Ur\ of the Stokes $\mathcal{Q}$ and $\mathcal{U}$ parameters, respectively \citep{schmid2006, millar-blanchaer2015}. The \Qr\ signal shares roughly the same extent and shape of the total intensity ring, with greater brightness along the front edge than the back edge. The inner hole is not prominent in \Qr, which suggests it may be enhanced by ADI self-subtraction in the total intensity images; this is supported by the modeling we show later (see Figure \ref{fig:modstack}). The southeast (SE) side of the disk appears brighter than the northwest (NW) side, particularly in the region where we assume the back edge to be. The \Ur\ image contains no clear disk signal but shows a quadrupole pattern that may result from instrumental polarization unsubtracted during reduction. We use this \Ur\ image to estimate noise in the \Qr\ data because we do not expect single scattering by circumstellar grains to generate a significant \Ur\ signal \citep{canovas2015}. Dividing the \Qr\ image by a noise map, built from the standard deviation of \Ur\ pixel values in 1-px wide annuli centered on the star, we created the \Qr\ signal-to-noise ratio (S/N) image shown in the right-hand panel of Figure \ref{fig:polstack}.

The ring's edges appear softer in \Qr\ than in total intensity, particularly on the SE side. This is likely because only the total intensity ring is biased by ADI self-subtraction, which partially resembles a high-pass filter. Soft edges might indicate a vertically extended ring. In this case, light scattered by the front edge may be conflated with light scattered by the back edge for scattering angles $>90\degr$. This would affect both polarized and total intensity measurements. We discuss this possibility further in Section \ref{sect:mod_phase}.

% Disk Properties
\subsection{Scattering Phase Functions} \label{sect:phase_funcs}

\begin{figure}[h!]
\centering
\includegraphics[width=\columnwidth]{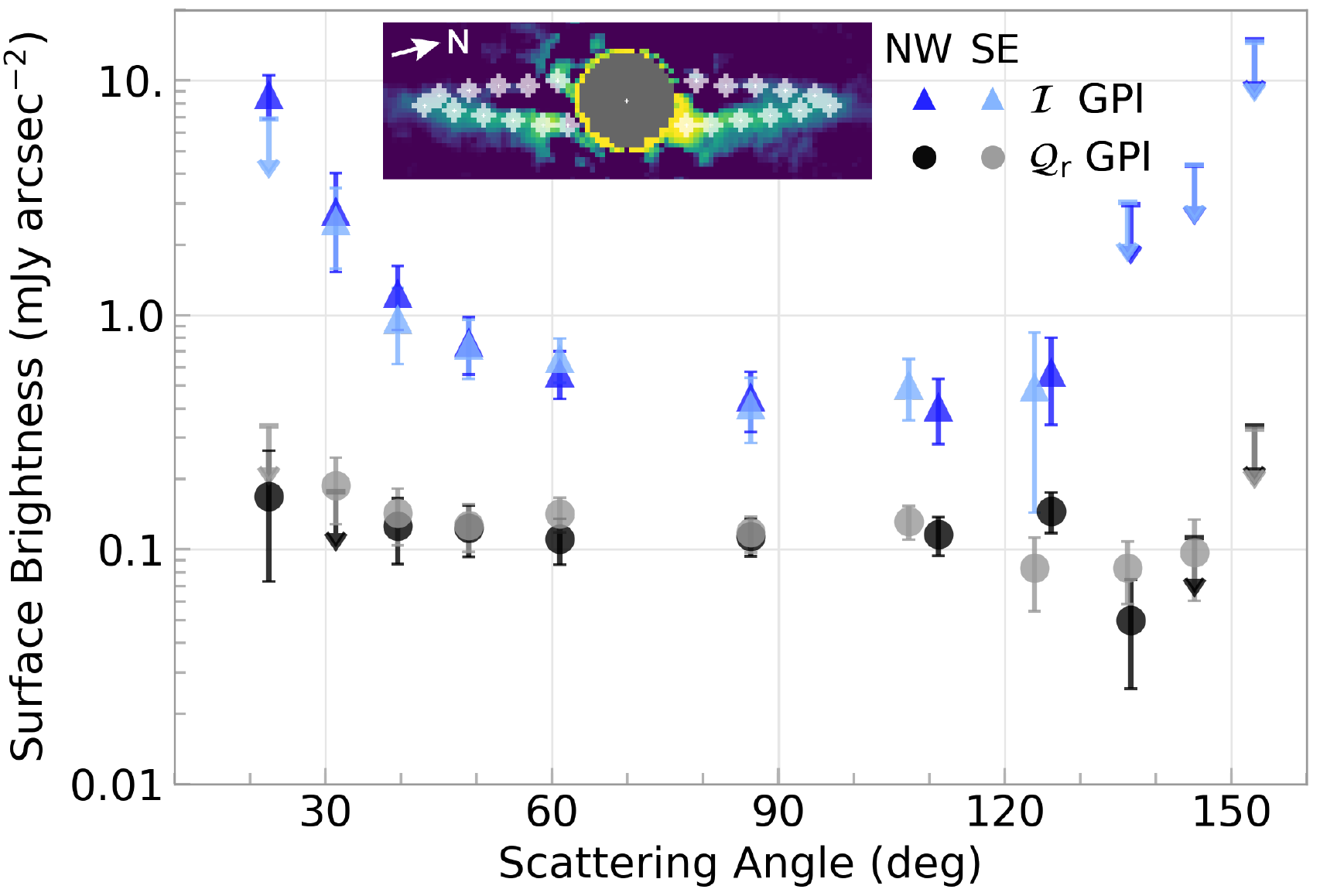}
\caption{Ring surface brightness in GPI $H$-band total intensity (blue) and Stokes $Q_\mathrm{r}$ (gray) as a function of scattering phase angle. The profiles are divided into the northwest and southeast sides of the ring. Brightness values have been corrected for ADI self-subtraction bias via forward-modeling. Errors are 1$\sigma$ uncertainties and arrows without markers are 3$\sigma$ upper limits with arrow lengths of 1$\sigma$. The inset shows a map of the apertures used.}
\label{fig:bright_prof}
\end{figure}

We quantitatively assessed the disk's scattering phase function by measuring its surface brightness as a function of scattering angle $\theta$ (Figure \ref{fig:bright_prof}). These angles assume a circular ring centered on the star with an inclination of $84\fdg9$ (determined from modeling described in Section \ref{sect:modeling}). To measure the scattering phase function, we placed apertures (radius = 2 px) on the conservative KLIP ring at a range of scattering angles from 22$\degr$, located closest to the star on the front edge, to 154$\degr$, closest to the star on the back edge (see Figure \ref{fig:bright_prof} inset). The ansae are at $\theta\approx90\degr$. The NW and SE sides of the ring were measured independently.

To estimate the self-subtraction of disk brightness by KLIP PSF subtraction, we forward modeled the effect with the ``DiskFM'' feature included in \texttt{pyKLIP}. This projects the relevant principal components onto a model of the disk's underlying brightness distribution and considers the effects of disk signal leaking into the principal components, accounting for both over- and self-subtraction of the disk. For the underlying brightness, we used the ``median model'' result from our MCMC fit, described in Section \ref{sect:model_results}. We then computed the ratio of that underlying brightness to the corresponding forward-modeled brightness at every pixel to get a correction factor for each aperture, and finally multiplied each aperture measurement by said factor to get the corrected brightness values plotted in Figure \ref{fig:bright_prof}. All of our total intensity brightness and flux measurements include these corrections.

The error bars represent 1-$\sigma$ uncertainties. For each data point, they are the quadrature sum of (1) the standard deviation of mean surface brightnesses within apertures placed at the same separation but outside the disk and (2) an assumed 15\% error on the self-subtraction correction factor based on variances of measurements for different models and reductions. For measurements that are consistent with zero brightness according to our uncertainties, we plot 3-$\sigma$ upper limits only.

% Scattering Phase Function.

We find the ring's brightness to be symmetric with scattering angle between its SE and NW halves. The one exception is our innermost measurement at $\theta\approx22\degr$, for which our errors may be underestimated due to non-Gaussian noise from residual speckles close to the star. The brightness along the ring's front edge decreases by a factor of ${\sim}20$ from $\theta\approx22\degr$ to the ansae. The ring brightness along the back edge ($\theta>90\degr$) is roughly consistent with being constant in $\theta$, although it is largely unconstrained at $\theta>125\degr$. This general shape is consistent with several other debris disks with measured phase functions \citep{hughes2018_review}.

We performed similar brightness measurements on the \Qr\ image, also plotted in Figure \ref{fig:bright_prof}. The uncertainties are calculated from the \Ur\ image as the standard deviation of mean surface brightnesses within apertures placed at the same separations as the data measurements. This assumes the noise properties are similar between \Qr\ and \Ur. No self-subtraction corrections are needed for our polarized intensities.

There is less variation of the polarized intensity with $\theta$ than for total intensity, as the front edge is only about 1.5--2.0 times brighter than the ansae. The back edge brightness again has relatively large uncertainties, however, it may be slightly fainter than the ansae. The SE side of the ring is preferentially brighter than the NE side, particularly on the front edge, but the asymmetry is marginal given our photometric precision. Nevertheless, these phase functions provide useful points of comparison for our models, particularly regarding grain properties.

\subsection{Polarization Fraction}

\begin{figure}[ht!]
\centering
\includegraphics[width=\columnwidth]{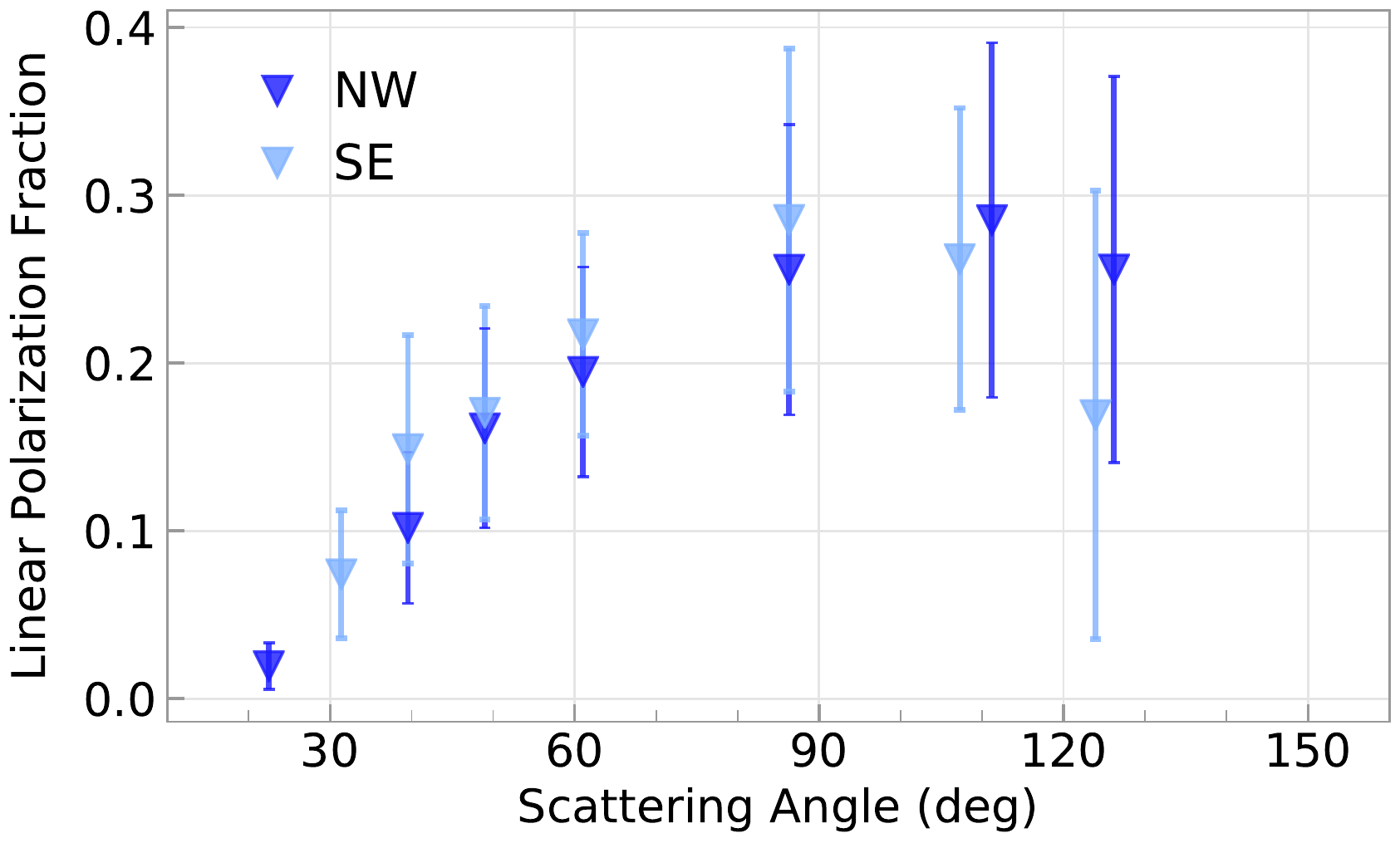}
\caption{The ring's polarization fraction as a function of scattering phase angle. The northwest and southeast sides of the ring are plotted separately. Points are not plotted for which we have only upper limits on the total intensity or \Qr\ brightness.}
\label{fig:pol_frac}
\end{figure}

Having brightness measurements for both total and polarized intensity, we computed their ratio to get a polarization fraction for the ring, plotted in Fig~\ref{fig:pol_frac}. The 1-$\sigma$ uncertainties were propagated from the uncertainties on both sets of brightness measurements and we exclude measurements for which we have only upper limits on the total intensity or Qr brightness.

The polarization fraction is generally higher to the SE than the NW but not to a significant degree given our uncertainties. The fraction peaks near the ansae at ${\sim}25$--30\% but may be as low as a few percent at the smallest scattering angles. The location of the peak near $\theta=90\degr$ (our measurement is at $\theta=87\degr$) is consistent with most predictions for scattering by micron-sized grains. Large uncertainties on brightness measurements along the back edge make for poorly constrained polarization fractions at large scattering angles; however, we see a tentative trend of the fraction decreasing as the angle increases past $90\degr$. We do not report the fraction for $\theta>130^\circ$ because it is unconstrained between 0\% and 100\% within the 3$\sigma$ uncertainties on our total intensity and \Qr\ brightness measurements.

% Polarization fraction.

The HD 35841 polarization fractions are similar to those measured for other debris disks. For example, AU Mic \citep{graham2007} and HD 111520 \citep{draper2016} both peak at 40\% at the largest separations, which may be the ansae of those edge-on rings.

\subsection{Disk Color}

To compute a STIS$-$GPI $H$ color, we first measured the flux within one 3x3 px aperture centered on each ansa in the GPI conservative KLIP image and the interpolated STIS image. We make this measurement at the ansae because they are the only places that we detect the disk with both GPI and STIS. The same aperture centers were used for both images, with locations relative to the star of ($r$,PA) = ($0\farcs58$,$165\fdg8$) and ($0\farcs58$,$-14\fdg2$) for the SE and NW ansae, respectively. Both correspond to a scattering angle of $87\degr$. The NW aperture lies just outside of the STIS image's masked region. We then subtracted a stellar STIS$-$GPI $H$ color of 1.10 mag from the difference of the fluxes. The stellar color is based on the 2MASS $H$ magnitude and an $8.88\pm0.01$ mag in the STIS 50CCD bandpass (converted from the Tycho 2 $V$-band value of $8.90\pm0.01$ mag; \citealt{hog2000}).

We measure the STIS$-$GPI $H$ color to be $-0.23\substack{+0.09 \\ -0.05}$ mag and $-0.26\substack{+0.09 \\ -0.05}$ mag for the ring's SE and NW ansae, respectively. This makes the ring slightly blue on both sides, along the lines of the optical vs. near-IR colors of debris disks like AU Mic (\citealt{lomax2017} and references therein) and HD 15115 (which is blue in $V-H$ according to \citealt{kalas2007a} and \citealt{debes2008}). With only two measurements, we limit our speculation as to the physical interpretation of the disk color and look forward to future visible-light observations that resolve more of the ring.

%\pagebreak % for latex arrangement
% --- PLANET SENSITIVITY --- %
\subsection{Point-Source Sensitivity} \label{sect:planets}

Our observations yielded no significant point sources. Based on our data, we determined limits on our sensitivity to substellar companions around HD 35841. For this, we only consider H-Spec because it achieved deeper contrast than H-Pol and lower mass limits than STIS.

We based our contrast measurements on reductions optimized for point-source detection and separate from those already presented. In this case, we used \texttt{pyKLIP} on the full 44-image data set, taking advantage of both angular and spectral diversity (i.e., ADI + SDI). Images were divided into 9 equal-width annuli between $r=10$ and 115 px that were split azimuthally into four subsections. References were restricted by a minimum rotation threshold of 1 px. We selected the first 30 KL modes among the 300 most correlated references for each target image (more references are available now because we do not spectrally collapse the data). The PSF was subtracted assuming two different spectral templates for a hypothetical planet: one with a flat spectrum and one with a methane-absorption spectrum (e.g., similar to that of 51 Eri b; \citealt{macintosh2015}). To correct for point source attenuation by the KLIP algorithm, we injected fake Gaussian point sources into the input images and then recovered their fluxes after PSF subtraction. The fake planet spectrum was matched to the reduction type, as either flat or methane-absorbing.

% Contrast.
\begin{figure}[h]
\centering
\includegraphics[width=\columnwidth]{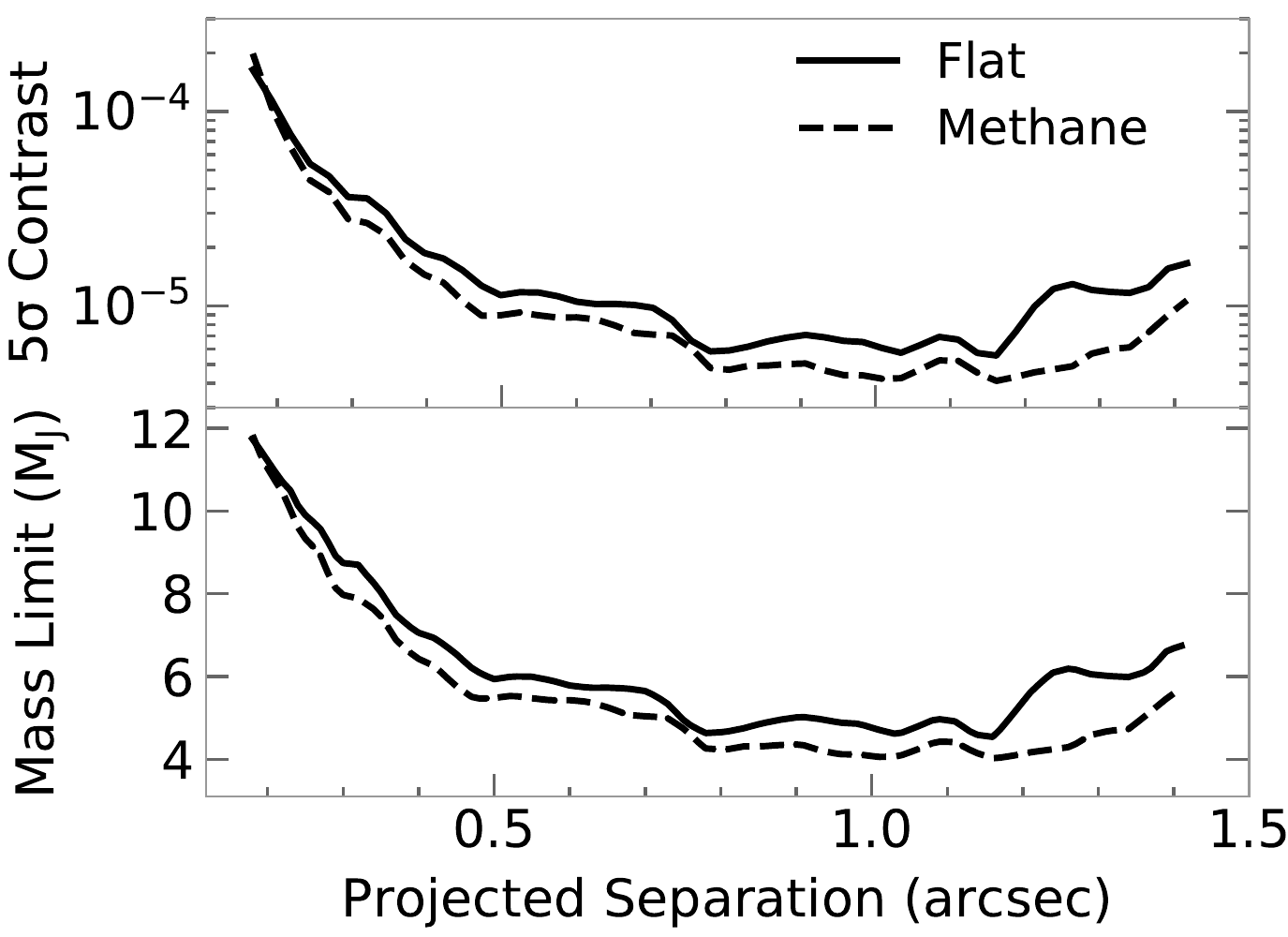}
\caption{Top: The 5$\sigma$ contrast limits from our H-Spec data, assuming either a flat or methane-absorption planet spectrum. Bottom: The contrasts are converted to mass limits for ``hot start'' planets.}
\label{fig:contrast}
\end{figure}

Our 5$\sigma$ equivalent point-source contrast limits \citep{mawet2014, wang2015_aumic}, corrected for PSF subtraction throughput, are shown in Figure \ref{fig:contrast}. We translated these contrast values into planet mass limits using AMES-Cond atmosphere models \citep{allard2001, baraffe2003} to convert planet luminosity to mass assuming an age of 40 Myr and ``hot start'' formation. At moderate separations of $0\farcs8$--$1\farcs3$ (82--134 au) we can rule out planets more massive than 5 $\mathrm{M_J}$ and can more generally exclude 12 $\mathrm{M_J}$ companions or larger at $0\farcs2$--$1\farcs4$ (21--144 au). These limits only apply to regions outside of the ring, however, as a planet embedded in or interior to the ring could be obscured by the dust. Therefore, we are most sensitive to companions that are not coplanar with the disk.

% --- MODELING SECTION --- %
\section{DISK MODELING} \label{sect:modeling}

We made a variety of comparisons between the data and models to explore possible disk compositions and morphologies. All of the models were constructed using the MCFOST radiative transfer code \citep{pinte2006} and employed spherical grains affecting Mie scattering in an optically thin disk. To fit these models to data we employed Markov-Chain Monte Carlo (MCMC) simulations using the Python module \texttt{emcee} \citep{foreman-mackey2013}.

% --- MODEL DESCRIPTION --- %
\subsection{The Debris Disk Model}

Our underlying disk model distributes grains in an azimuthally symmetric ring that is centered on the star. It consists of one component, which \citet{soummer2014} found sufficient to fit the disk's SED. We chose MCFOST's ``debris disk'' option to define the disk's volume density profile, which follows the form

% Disk density distribution.
\begin{equation}
%\begin{aligned}
\rho(r,z) \propto \frac{\exp\big\{-\big[|z|/H(r)\big]^\gamma\big\}}{\big[\left(r/R_c\right)^{-2\alpha_{in}}+\left(r/R_c\right)^{-2\alpha_{out}}\big]^{1/2}}\ , \\
\label{eq:disk_density} \\
%\end{aligned}
\end{equation}

\noindent where $r$ is the radial coordinate in the equatorial plane, $z$ is the height above the disk midplane, and $R_c$ is a critical radius that divides the ring into inner and outer regions with density power law indices of $\alpha_{in}$ and $\alpha_{out}$, respectively \citep{augereau1999}. The disk scale height varies as $H(r) = H_0(r/R_0)$ such that the scale height is $H_0$ at radius $R_0$, while the slope of the vertical density profile is constrained by the exponent $\gamma$. We chose to set $R_0=60$ au because that radius appears by eye to pass through the middle of the ansae, i.e., it is the midpoint between the ring's inner and outer edges.

The ring's inner and outer edges are set by $R_{in}$ and $R_{out}$, respectively, and tapered by a Gaussian with $\sigma=2$ au so the volume density smoothly declines to zero (separate from Eq. \ref{eq:disk_density}). We found the outer radius to be weakly constrained in preliminary small-scale MCMC tests because the ring's edge gradually blends into the GPI data's background noise at ${\sim}80$ au. However, we know from the STIS image that the disk extends out to at least 110 au. Therefore, we set a conservative outer ring radius of twice this distance, $R_{out}=220$ au, and hold it constant throughout the fitting process. The viewing geometry of the ring is set by the disk inclination $i$ and position angle $PA$.

We populate the single disk component with grain particles following the power-law size distribution $\mathrm{dN(}a\mathrm{)}\propto a^{-q}\ \mathrm{d}a$, where the grain size $a$ ranges from a minimum size $a_{min}$ to a maximum size of 1 mm. We consider grains composed of three different materials\footnote{The MCFOST optical indices are derived from the following sources: amorphous Si similar to \citet{draine1984_si}, aC from \citet{rouleau1991_carbon}, and H$_2$O compiled from sources described in \citet{li1998_ice}}: astrosilicates (Si), amorphous carbon (aC), and water ice (H$_2$O). The relative abundances of these compositions are defined as fractions of the total disk mass (\msi, \mac, \mice) and are allowed to vary so long as their mass fractions sum to 1. However, all grains have pure compositions (e.g., no individual grain is 50\% Si, 50\% aC), are distributed in the same manner throughout the disk volume regardless of composition, and share the same size distribution and porosity within a given model. MCFOST approximates porous particles by ``mixing'' a grain's material composition with void (refractive index of $n=1+0i$); the mixture is performed using the so-called Bruggeman mixing rule of effective medium theory to get the effective refractive index of the grains. The total dust mass $M_d$ regulates the model's scattered-light surface brightness and thermal flux.

We do not include radiation pressure, Poynting-Robinson (P-R) drag, or gas drag effects in our model. Being relatively bright, we expect this disk's dust density to be high enough that P-R drag is negligible \citep{wyatt2005a}. Only a non-detection of gas has been reported for this disk \citep{moor2011b}, so we assume a standard debris disk scenario in which most of the protoplanetary gas has been removed and gas drag is also negligible. We exclude radiation pressure for simplicity, but it may have important effects on the outer disk, which we discuss later.

\subsection{MCMC Modeling Procedure}

We derived the disk's primary morphological and grain parameters by fitting scattered-light models to our GPI total intensity and \Qr\ images via MCMC. Only the LOCI image of the total intensity was used in the MCMC because it allowed for faster forward modeling of the ADI self-subtraction than did the KLIP image. The STIS image was not included in the fit due to its limited coverage of the ring but was used for comparison afterwards.

Uncertainty maps, calculated as the standard deviation among pixels in the data at the same projected radius from the star, were constructed at the original GPI resolution for both total intensity and \Qr. We then binned the images and uncertainty maps into 2$\times$2 px bins to mitigate correlation between pixels within the same resolution element ($\sim$3.5 px across) in our data. Ideally to remove correlations, the bin size should be equal to the size of one resolution element, but we found that binning the data to this degree removed many of the disk's defining morphological features. Alternatively, the correlations between different elements can be measured and explicitly taken into account as part of the fitting process \citep{czekala2015b, wolff2017b}.

For the $H$-band models, we scattered only photons with a wavelength of 1.647 $\micron$, approximate to the central wavelength of the GPI $H$ bandpass. We found use of a single wavelength to be a computationally efficient proxy for integrating models over the entire bandpass. To compare models to data in each iteration of the MCMC, we first constructed the models at the same resolution as the original GPI data. We then convolved them with a normalized 2-d Gaussian function with a 3.8-px full-width half-maximum to approximate the GPI PSF. For total intensity only, we then forward modeled the LOCI self-subtraction using the ``raw'' total intensity model output by MCFOST as the underlying brightness distribution \citep{esposito2014, esposito2016}. It is this forward-modeled version that we compare with the LOCI image, providing a fair comparison to the self-subtracted data. No forward modeling was necessary for the \Qr\ models.

Our parallel-tempered MCMC used three temperatures with 150 walkers each. We initialized walkers randomly from a uniform distribution, and then simulated each walker for 11,000 iterations (\num{4.95e6} samples). Parameter values were constrained by a flat prior with the limits listed in Table \ref{tab:bf_params}. Ultimately, the walker chains evolved significantly for ${\sim}$10,000 iterations before stabilizing (i.e. ``converging''), therefore, our posterior distributions are drawn from the final 1,000 iterations (\num{1.5e5} samples) of the zeroth temperature walkers only.

% --- MCMC Results --- %
\subsection{MCMC Modeling Results} \label{sect:model_results}

The results of the MCMC are listed in Table \ref{tab:bf_params} as the parameter values associated with the maximum likelihood model (i.e., ``best fit'') and also the 16th, 50th (i.e. median), and 84th percentiles of the marginalized posterior probability distribution functions (PDF). A corner plot for those distributions is provided in Appendix \ref{sect:appA}.

% --- BEST-FIT MODEL PARAMETER TABLE --- %
\begin{table}[h]
\begin{center}
\caption{MCMC Model Parameters}
\label{tab:bf_params}
\begin{tabular}{l c c c c c} %l}
\toprule
Param. & Limits & Max Lk & 16\% & 50\% & 84\% \\ %& Unit  \\ % & Description  \\
\midrule

% Numbers.
$i$ (deg) & [80, 88] & 85.1 & 84.7 & 84.9 & 85.1 \\ %& deg \\ %& Inclination
\textit{PA} (deg) & [163, 167] & 165.9 & 165.6 & 165.8 & 165.9 \\ %& deg \\ %& on-sky PA
$R_{in}$ (au) & (10, 65] & 59.9 & 57.7 & 59.8 & 60.9 \\ %& au \\ %& Inner radius
$H_0$ (au) & (0.3, 10] & 2.4 & 2.4 & 2.7 & 4.1 \\ %& au \\ %& Scale height
$\gamma$ & (0.1, 3] & 0.52 & 0.51 & 0.56 & 0.68 \\ %&  \\ %& Vertical density exp
$M_d$ (M$_{\oplus}$) & (0.00053, 3.3) & 0.11 & 0.11 & 0.15 & 0.19 \\ %& Dust mass
$a_{min}$ ($\micron$) & [0.1, 40] & 1.5 & 0.16 & 1.5 & 1.6 \\ %& $\micron$ \\ %& Min grain size
$q$ & (2, 6) & 2.9 & 2.7 & 2.9 & 3.0 \\ %&  \\ %& Grain size dist. exponent
\mac & (0, 1) & 0.63 & 0.35 & 0.48 & 0.61 \\ %& \\ %& Mass frac carbon
\mice & (0, 1) & 0.31 & 0.21 & 0.27 & 0.33 \\ %&  \\ %& Mass frac water ice
$\alpha_{out}$ & (-6, 0) & -2.9 & -3.2 & -3.0 & -2.8 \\ %&  \\ %& Outer vol dens exp ('gamma_exp')
\multicolumn{6}{c}{Parameters considered upper or lower limits*}
\\
$\alpha_{in}$ & (-6, 7) & 4.0 & -1.6 & 3.8 & 6.2 \\ %&  \\ %& Inner vol dens exp ('surface_density_exp')
$R_{c}$ (au) & (40, 110] & 51.0 & 45.8 & 53.2 & 57.0 \\ %& au \\ %& Critical radius
porosity (\%) & (0, 95) & 1.4 & 0.50 & 1.5 & 3.3 \\ %&  \\ %& Porosity
\msi & (0, 1) & 0.058 & 0.079 & 0.24 & 0.42 \\ %&  \\ %& Mass frac silicates

\bottomrule
\end{tabular}
\end{center}
* The bottom four parameters all have posterior PDF's that extend to either an upper or lower boundary imposed by our MCMC prior. Therefore, these should be formally considered lower ($\alpha_{in}$) or upper limits ($R_{c}$, porosity, \msi).
\end{table}

We find the ring to be inclined 84$\fdg9^{+0.2}_{-0.2}$ and ${\sim}160^{+1.1}_{-2.1}$ au wide, with dust between radii of $59.8^{+1.1}_{-2.1}$ au and 220 au. Vertically, the scale height is $2.7^{+1.4}_{-0.3}$ au at $R_0$, with a density profile exponent less than unity ($\gamma=$ 0.51--0.68). Both parameters are weakly bimodal, favoring vertically thin disks ($H_0\approx2.4$ au, $\gamma\approx0.52$) but also showing thicker disks ($H_0\approx4.1$ au, $\gamma\approx0.68$) to agree nearly as well with the data. The marginalized PDF for $R_c$ abuts our lower prior boundary of 40 au, so we only place an upper limit of 57.0 au (its 84th percentile value) on it. However, $R_c < R_{in}$ suggests a sharp inner edge to the ring. This also makes $\alpha_{in}$ degenerate in most cases, so we can only place a lower limit for it at $-1.6$ (its 16th percentile value). The outer volume density power law $\alpha_{out}$ is better defined at $-3.0^{+0.2}_{-0.2}$. Therefore, the dust density decreases continuously from a peak near $R_{in}$ to the outer edge. The \textit{PA} is $165\fdg8^{+0.1}_{-0.2}$.

% Model and Residual images.
\begin{figure*}[ht]
\centering
\includegraphics[width=6.5in]{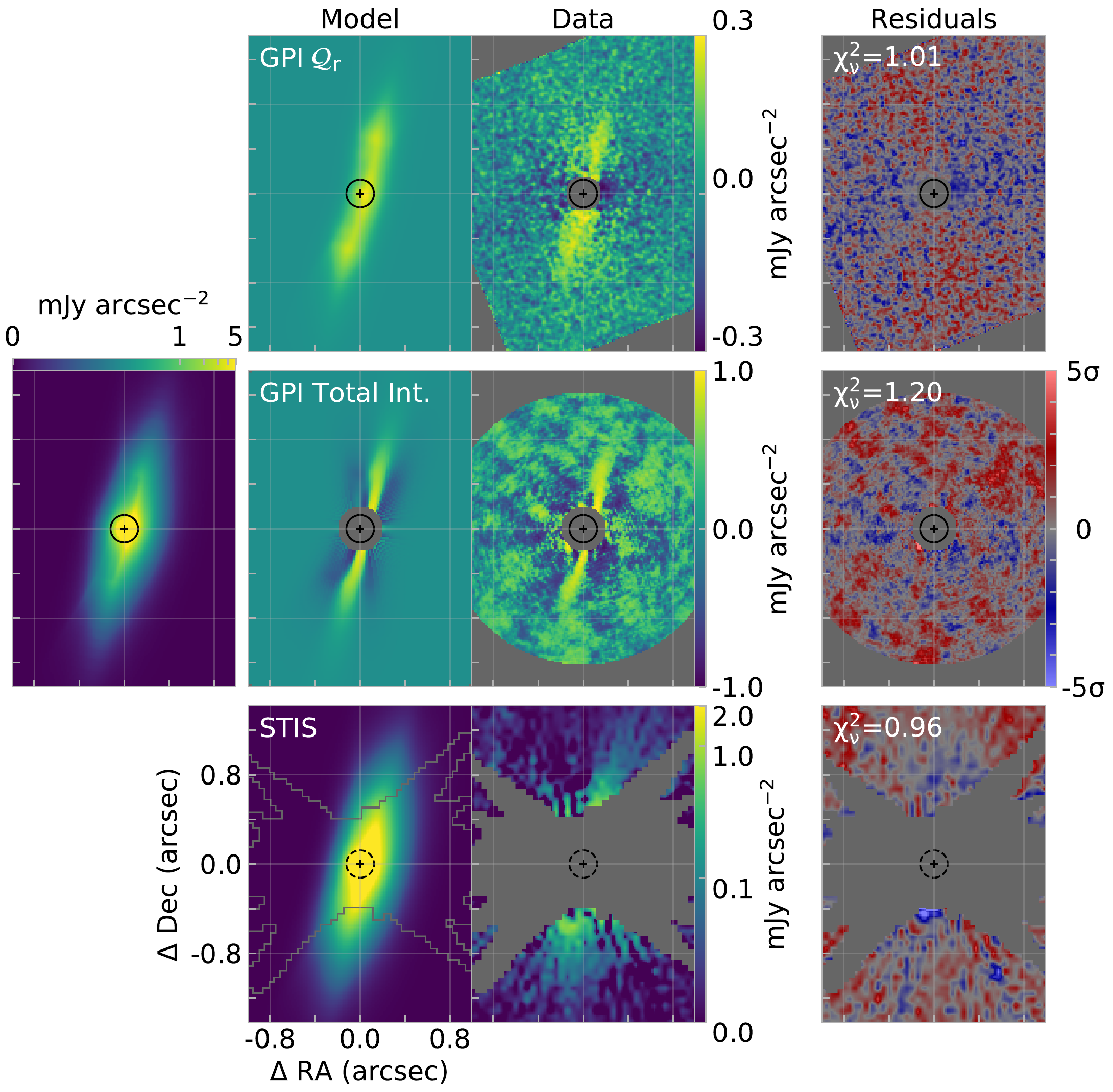}
\caption{Models constructed from median MCMC parameter values compared with our GPI Stokes \Qr\ (\textit{top}), total intensity (\textit{middle}), and STIS data (\textit{bottom}). The left panel in the middle row shows the $H$ total intensity model without ADI forward-modeling. The data and models have a log color scale and the residuals have a linear scale. Dark gray regions mask areas in which we have no useful information due to high noise or masks. The black cross and circle mark the star and size of the GPI FPM, respectively.}
\label{fig:modstack}
\end{figure*}

The dust properties are less constrained than the ring's morphological properties. We find the most likely minimum grain size $a_{min}$ to be 1.5 $\micron$ but there is a weaker secondary peak in the marginalized posterior at ${\sim}0.16$ $\micron$. The blowout size ($a_{blow}$) for this star\footnote{Blowout size depends on the following assumed properties: $M_{*} =$ 1.29--1.31 M$_\odot$, $L_{*} =$ 2.4--3.1 L$_\odot$, grain density = 2.7 g cm$^{-3}$, and a radiation pressure efficiency $Q = 2$.} is ${\sim}$1.6--2.1 $\micron$; grains smaller than $a_{blow}$ experience a radiation pressure force greater than the star's gravitational force and are thus ejected from the system. This additional constraint leads us to accept the larger $a_{min}$ of ${\sim}1.5$ $\micron$ as the most likely value. The power law index of the grain size distribution is $2.9^{+0.1}_{-0.2}$ and the dust mass in grains with sizes between $a_{min}$ and 1 mm is approximately 0.1--0.2 $\mathrm{M}_\oplus$. The median mass fractions among the three compositions are 24\% \msi, 48\% \mac, and 27\% \mice, although \msi\ extends to the lower prior boundary of 0\%. We also note that the maximum likelihood model (which is thinner vertically) favors carbon more strongly, with mass fractions of 6\% \msi, 63\% \mac, and 31\% \mice. Grains with low porosity are strongly preferred overall, with a 99.7\% confidence upper limit of $<12\%$. We discuss some of the implications of these dust properties in Sections \ref{sect:disc_grain_shape} and \ref{sect:disc_composition}.

We present ``median model'' images alongside the data and data$-$model residuals in Figure \ref{fig:modstack}. This median model is constructed using the median value of each parameter's PDF from the MCMC. As Table \ref{tab:bf_params} shows, the median values are generally close to those of the maximum likelihood model, making the two models nearly indistinguishable to the eye and essentially equal in terms of $\chi^2_\nu$ (to two decimal places for GPI data and only differing by 0.01 for the STIS comparison).

The \Qr\ model is a good match to the data, with little residual disk signal. A quadrupole pattern is apparent in the residual map, which is a sign of instrumental polarization that was not completely removed during data reduction. The model's \Ur\ signal is at least 100 times below the noise floor of our data and consistent with no disk signal. The forward-modeled total intensity agrees well with the data along the ring's front edge but the model is comparatively faint at the ansae and the back edge. Our model grains, therefore, are more forward-scattering than the real grains. The underlying total intensity model (far left panel) contains a much more vertically extended scattered-light distribution than the forward-modeled version, demonstrating how the ADI data reduction sharpens the ring's edges and generally suppresses its surface brightness.

The same underlying model recomputed for 0.575-$\micron$ scattered light (bottom of Figure \ref{fig:modstack}) agrees well with the STIS image out to projected separations of ${\sim}110$ au but fails to account for halo brightness at larger separations. This is evidenced by positive residuals NW and SW of the star (the positive residuals to the NE are localized noise and not disk signal). The discrepancy may be even more pronounced near the minimum of $\theta$ where our data are incomplete. We know our model contains dust at these large separations but more scattering is required to match the observed brightness.

Thus, we propose that this outer halo brightness is produced by an additional population of grains slightly larger than $a_{blow}$ that are produced in the ring and then pushed onto eccentric orbits with large apoapses by radiation pressure \citep{strubbe2006}. We do not include radiation pressure directly in our models, so we do not expect them to contain scattered light from such a population. For a consistency check, however, we can approximate this eccentric dust with a separate, manually-tuned model. Doing so, we find that a rudimentary model of a broad annulus at 220--450 au containing \num{1.7e-3} $\mathrm{M}_\oplus$ of grains with $a=1.5$--3.0 $\micron$, the same inclination and composition as the ring, and $H_0=r/10$, appears qualitatively similar to the outer halo in the STIS image. Its $H$-band brightness is also below the GPI data's sensitivity limits (for the data reductions in this work), consistent with it going undetected with GPI.

% --- MODEL PHASE FUNCTIONS --- %
\subsection{Model Phase Functions} \label{sect:mod_phase}

\begin{figure}[h!]
\centering
\includegraphics[width=\columnwidth]{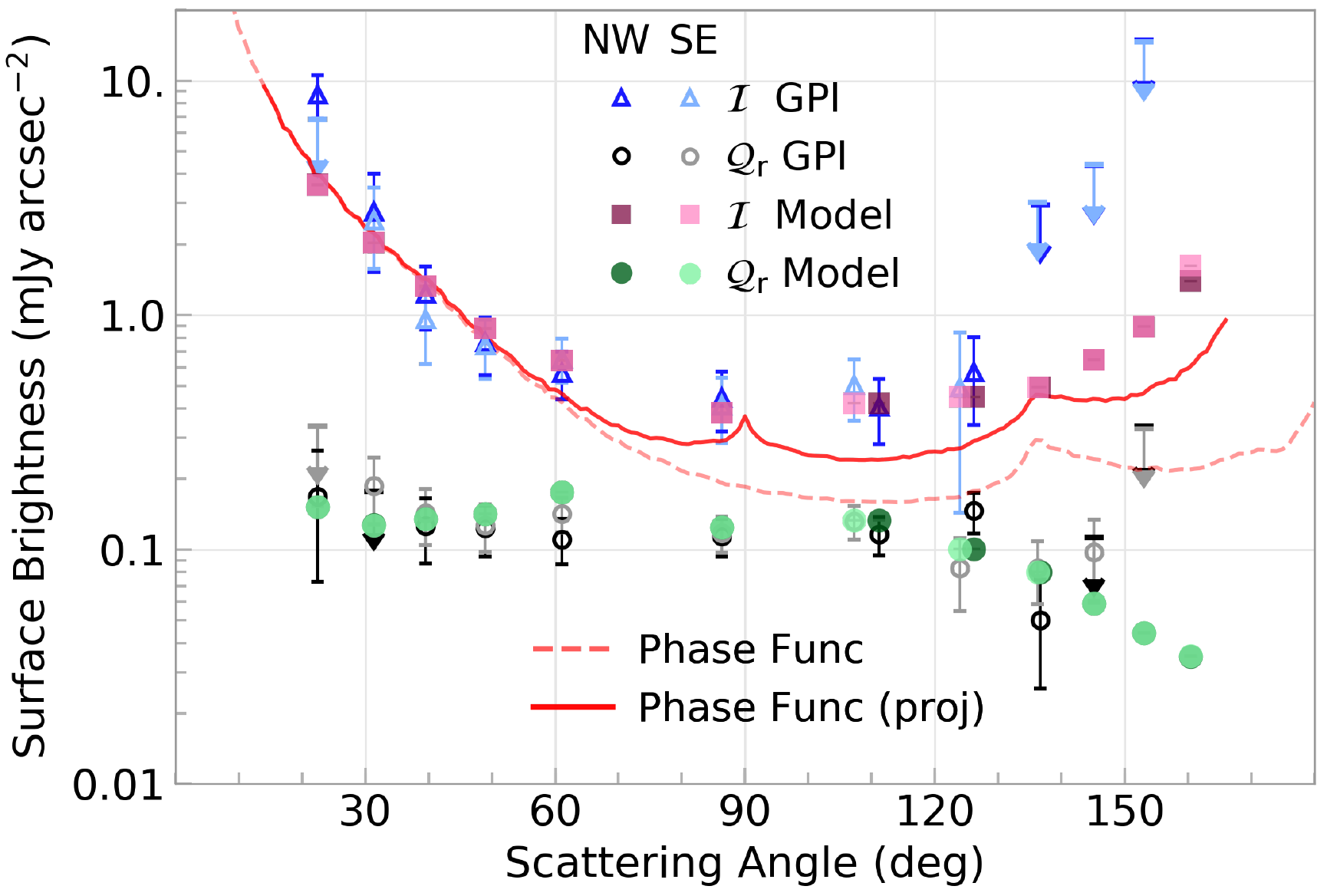}
\caption{Model scattering phase functions compared with our GPI-measured phase functions. The function directly implied by our scattered-light model and the data represents a projected phase function (solid red line), which is a standard phase function (dashed line) modified by the projection of some scattering angles onto others due to the disk shape and viewing geometry.}
\label{fig:mod_phasefuncs}
\end{figure}

We do not explicitly fit to the measured phase functions in Figures \ref{fig:bright_prof} and \ref{fig:pol_frac}; however, we do so implicitly when fitting the scattered-light images. Therefore, we can extract phase functions from our median model images, using the previously described aperture photometry method, and compare them with those measured from our data. Both are shown in Figure \ref{fig:mod_phasefuncs}. We find that the model's total intensity and \Qr\ brightnesses are generally consistent with observations at all measured scattering angles, with the best agreement coming at intermediate angles of 30$^\circ$--120$^\circ$.

% Model Phase Functions.

In addition to the aperture-measured profiles, we plot the analytic scattering phase function $B(\theta)$ for total intensity calculated by MCFOST for this model. The output is in arbitrary units, so we scaled it uniformly such that it equals the observed NW brightness point at PA=49$^\circ$. Comparing this analytic profile with the aperture profile for the same model, we find $B(\theta)$ to be consistently lower at $\theta\geq60^\circ$, i.e., the model's brightness is greater near the ansae and along the ring's back edge than expected, by more than 200\% at some angles. We believe this excess brightness results from light scattered by the front and back edges being conflated due to viewing the inclined and vertically thick ring in projection.

As a test of this hypothesis, we calculated a ``projected analytic phase function'' $B'(\theta)$ that takes into account scattered light from one edge being projected onto the other edge. We first estimated the vertical distance $\Delta z_j$ from the midplane of the ring at scattering angle $\theta_j$ to the midplane of the ring at the supplementary angle $180-\theta_j$ (i.e., the ``opposite'' edge). For all $\theta_j$, we then computed the fractional density of dust contributed from the supplementary scattering angle as $\Sigma_j = \exp{[(-|\Delta z_j|/H_0)^\gamma}]$, where $\Sigma=1$ at the supplementary midplane. The projected analytic phase function ends up as $B'(\theta_j)=B(180-\theta_j)\cdot\Sigma_j$, which we plot in Figure \ref{fig:mod_phasefuncs}. It is a closer match to the measured model phase function than the original analytic phase function is, though it still underestimates the scattering somewhat (by up to 50\% at $105^\circ < \theta < 125^\circ$). A localized peak occurs at $\theta=90^\circ$ where the projection effect is at maximum. A second peak at $\theta\approx136^\circ$ is produced by water ice preferentially scattering light incident at that angle, similar to the halo and sun dog effects seen for sunlight in Earth's atmosphere.

We conclude that the scattering phase function measured directly from disk photometry is significantly impacted by projection effects and should not be taken at face value as the pure phase function, particularly for highly inclined and vertically extended disks. It is especially important to take this into account when comparing analytic phase functions with empirical measurements.

% --- SED MODELS --- %
\subsection{Spectral Energy Distribution}

We modeled the disk's SED primarily as a check that our model's parameters were not ruled out by disk photometry at wavelengths outside of the near-IR. The data summarized in Table \ref{tab:phot} comprise a spectrum from the \emph{Spitzer Space Telescope} Infrared Spectrograph (IRS; \citealt{houck2004_irs}) and broadband photometric points from previous publications. This broadband photometry is composed of measurements from multiple optical and near-infrared instruments, NASA's Wide-field Infrared Survey Explorer (WISE; \citealt{wright2010_wise}), the Multiband Imaging Photometer for Spitzer (MIPS; \citealt{rieke2004_mips}), and the Submillimetre Common-User Bolometer Array 2 (SCUBA-2; \citealt{holland2013}).

% SED plot
\begin{figure}[h!]
\centering
\includegraphics[width=\columnwidth]{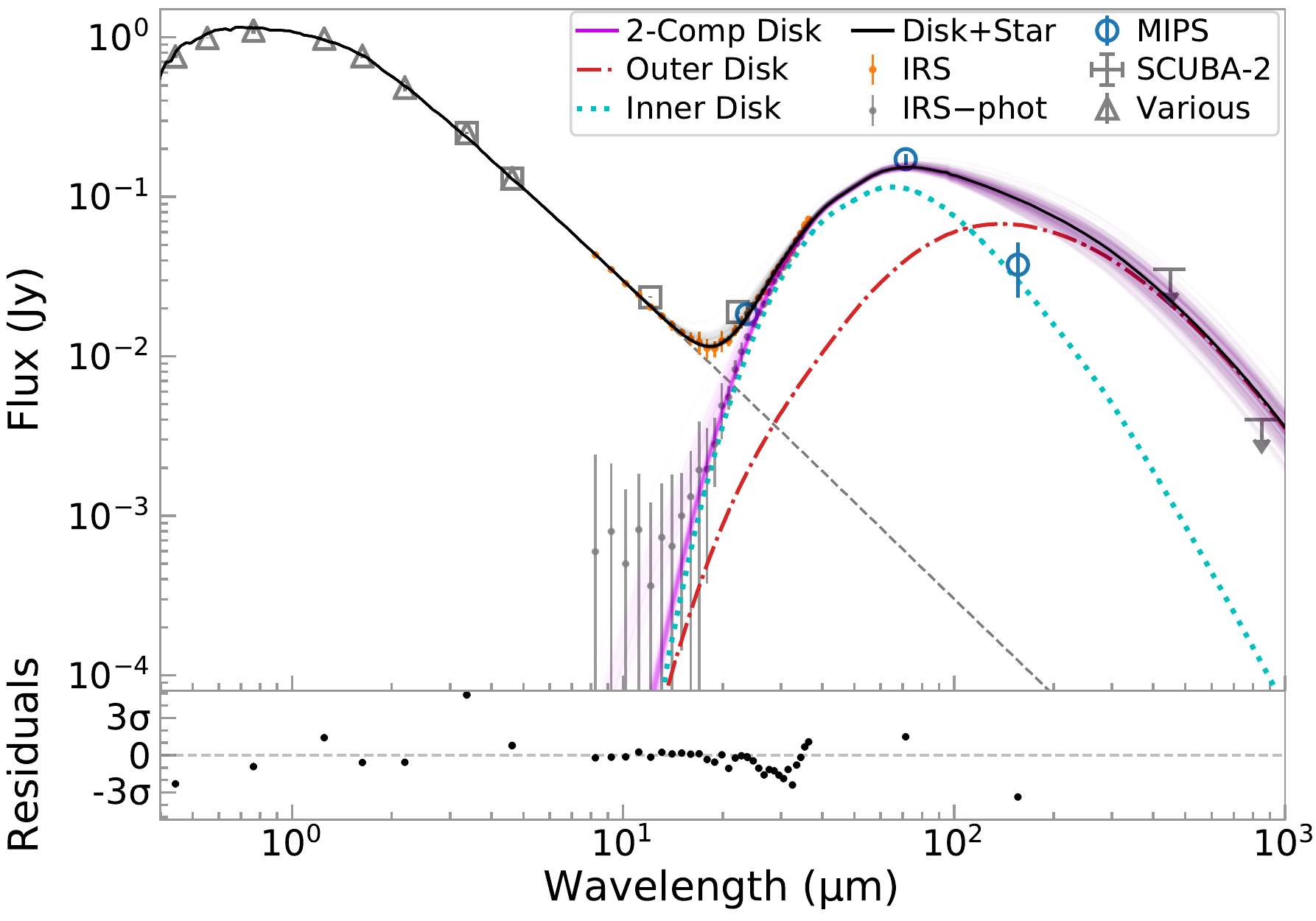}
\caption{A comparison of SED models with previously published photometry. The total SED for the MCMC median model (solid black line) is the sum of dust emission from that model (dot-dashed red), emission from a separate (manually-tuned) warm dust component (dotted cyan), and the stellar photosphere (dashed black). Residuals between the total median model and the data are in the bottom panel. We also plot 200 model SED's randomly drawn from the MCMC chain, with pink lines showing their dust emission only (MCMC model plus the warm component) and gray lines showing their dust emission plus the stellar photosphere. Orange points are the IRS spectrum binned into 1-$\micron$-wide bins. Gray points are the binned IRS spectrum with the stellar photosphere subtracted. The SCUBA-2 points at 450 $\micron$ and 850 $\micron$ are $5\sigma$ upper limits only.}
\label{fig:sed}
\end{figure}

In Figure \ref{fig:sed} we plot the data alongside the SED produced from the median model from our MCMC. The SED was computed with MCFOST, assuming a stellar spectrum model from \citet{kurucz1993} with effective temperature of 6460~K, stellar luminosity of 2.4~$L_\odot$, log surface gravity of 4.0, and solar metallicity, based on SED fitting results from \citet{moor2011a} and \citet{chen2014} updated to the Gaia-derived distance. This is the same stellar model used throughout our modeling efforts.

The median model SED is statistically consistent with the MIPS 160-$\micron$ point and upper limits at longer wavelengths but clearly deficient in flux at shorter wavelengths. This discrepancy may be due to another disk component not yet included in our model. This led us to model a separate inner disk component that, when added to the median model, would produce an SED similar to that observed. For simplicity, we manually tuned this inner component and consider it only a suggestion of one possible architecture for this disk.

Our best result assumes the inner component to be a narrow circular ring at $r=19$--20 au containing \num{2.1e-3} $\mathrm{M}_\oplus$ of dust. This inner component has the same inclination, fractional composition, and porosity as our median model but a larger minimum grain size of 10.0 $\micron$ and steeper size distribution with $q=4.5$. Being ${\sim}40$ au closer to the star and roughly 50 times less massive than the median model, this component does not significantly impact our scattered-light models and would be indistinguishable from noise in our observations.

For comparison with the measured SED, we randomly selected 200 models from the MCMC chains to serve as the outer components and added the inner component to each to produce a distribution of two-component SED's (Figure \ref{fig:sed}). We find that this distribution provides a good enough match to the data that our median ring model remains plausible given a suitable inner component. The two-component SED falls within 2--3 $\sigma$ of all measurements apart from exceeding the 160-$\micron$ MIPS flux by 3.5 $\sigma$ and the 850 $\micron$ upper limit by ${\sim}35$\%. The latter discrepancies stem from overproduction of flux by the median model at those wavelengths, which would be reduced if $R_{out}$ is smaller than our loosely assumed 220 au and there is less cold dust in the ring as a result.

% DISCUSSION
\section{DISCUSSION} \label{sect:discussion}

\subsection{Debris Disk Structure} \label{sect:disc_morph}

Our observations from ground- and space-based instruments, combined with MCMC modeling, have shown the HD 35841 system to include at least two, and perhaps three, debris disk components. Moving outward from the star, these are: an hypothetical inner dust component, a primary dust ring, and a smooth halo extending outward from the ring. This configuration shares similarities with many other stellar systems, including our own.

The $\sim$57--80 au spatial scale of the primary dust ring makes it nearly two times larger in radius than the Kuiper Belt (located at ${\sim}30$--50 au; \citealt{levison2008}). With HD 35841 being ${\sim}20$\% more massive than the Sun and possibly hosting a narrow inner component akin to an asteroid belt, this system resembles a scaled-up version of the Solar System. We find this ring's scale height to be 4\%--7\% of its stellocentric radius, which is in line with measurements of 3\%--10\% for other disks like HR 4796A, Fomalhaut, AU Mic, and $\beta$ Pic \citep{augereau1999, kalas2005_fomb, krist2005, millar-blanchaer2015}. Despite the HD 35841 ring not being exceptionally thick, the measured scattering phase function is still significantly impacted by projection effects because the ring is highly inclined. Therefore, we reiterate that this is an important aspect to consider for phase function measurements of other disks. This issue also highlights the value of polarimetry, which avoids PSF subtraction biases and enables additional constraints on disk geometry and phase functions.

A comparison of our models with the observed SED implies the existence of a second dust component interior to the ring imaged by GPI and STIS. Our brief exploration of solutions shows this inner component could be a narrow ring at ${\sim}20$ au ($0\farcs2$), which is just interior to our high SNR region in the GPI data but still outside of GPI's $H$-band inner working angle. It contains less mass than the main ring but it will receive more incident flux, buoying the scattered-light brightness. Future direct imaging with a smaller inner working angle may resolve this dust. Interferometric observations may also help constrain it.

We note that our disk components differ from those of previous SED fits of HD 35841. \citet{chen2014} fit blackbody rings at separations of 45 and 172 au (for a stellar distance of 96 pc) to the IRS and MIPS data. \citet{moor2011a} used just a single infinitesimally narrow ring of modified blackbodies at ${\sim}23$ au (also for $d=96$ pc). However, neither of those studies benefited from spatially resolved imaging, which requires dust at 57--80 au. Regarding their placement of dust interior to our primary ring, it is common to measure a resolved disk radius that is greater than the radius predicted by blackbody approximations, as \citet{morales2016} demonstrated for \emph{Herschel}-resolved disks. On the other hand, the outer component from \citet{chen2014} is located just beyond the outer part of the detected halo and could represent that material.

One final morphological feature of the disk not represented in our MCMC models is the outermost part of the smooth halo extending from the ring in the STIS image. Similar features are seen in other disk images, e.g., of HD 32297, HD 61005, and HD 129590 \citep{schneider2014, matthews2017}. These halos may be populated by ring grains that are slightly larger than $a_{blow}$ and are excited onto highly eccentric orbits by radiation pressure \citep{strubbe2006}. This type of eccentric grain population would cover a large surface area and scatter substantial light, but contain little mass and be relatively far from the star; thus, it would contribute little to the overall SED. This is qualitatively similar to what we find for HD 35841. One could better test the link between the ring and halo populations with more holistic models that directly incorporate radiation pressure into the model physics. However, theory predicts that the bound grain population should also leave an observational signature in the halo's surface brightness radial profile, which we test below.

\subsection{Bound Grains in the Halo} \label{sect:halo}

We measured surface brightness radial profiles for the smooth halo to see if their slopes agree with that expected for the bound, eccentric population of grains we proposed in the previous section.

To measure the radial profile, we placed apertures (radii = 2 px) along the ring's presumed major axis in the interpolated STIS image. The innermost apertures were centered at $r=51$ px (74.3 au), just exterior to the ansae, and were located every 5 px out to $r=237$ px (346 au) on both sides of the star. Surface brightnesses and their uncertainties were measured in the same way as the phase functions in Section \ref{sect:phase_funcs}. Measurements consistent with zero at less than the $1\sigma$ level are discarded, which includes all points at $r>210$ au. We then fit power-law functions of the form $\mathrm{SB}\propto r^{\alpha_h}$ independently to the radial profiles for each side of the halo using a least-squares minimizer algorithm.

\begin{figure}[h!]
\centering
\includegraphics[width=\columnwidth]{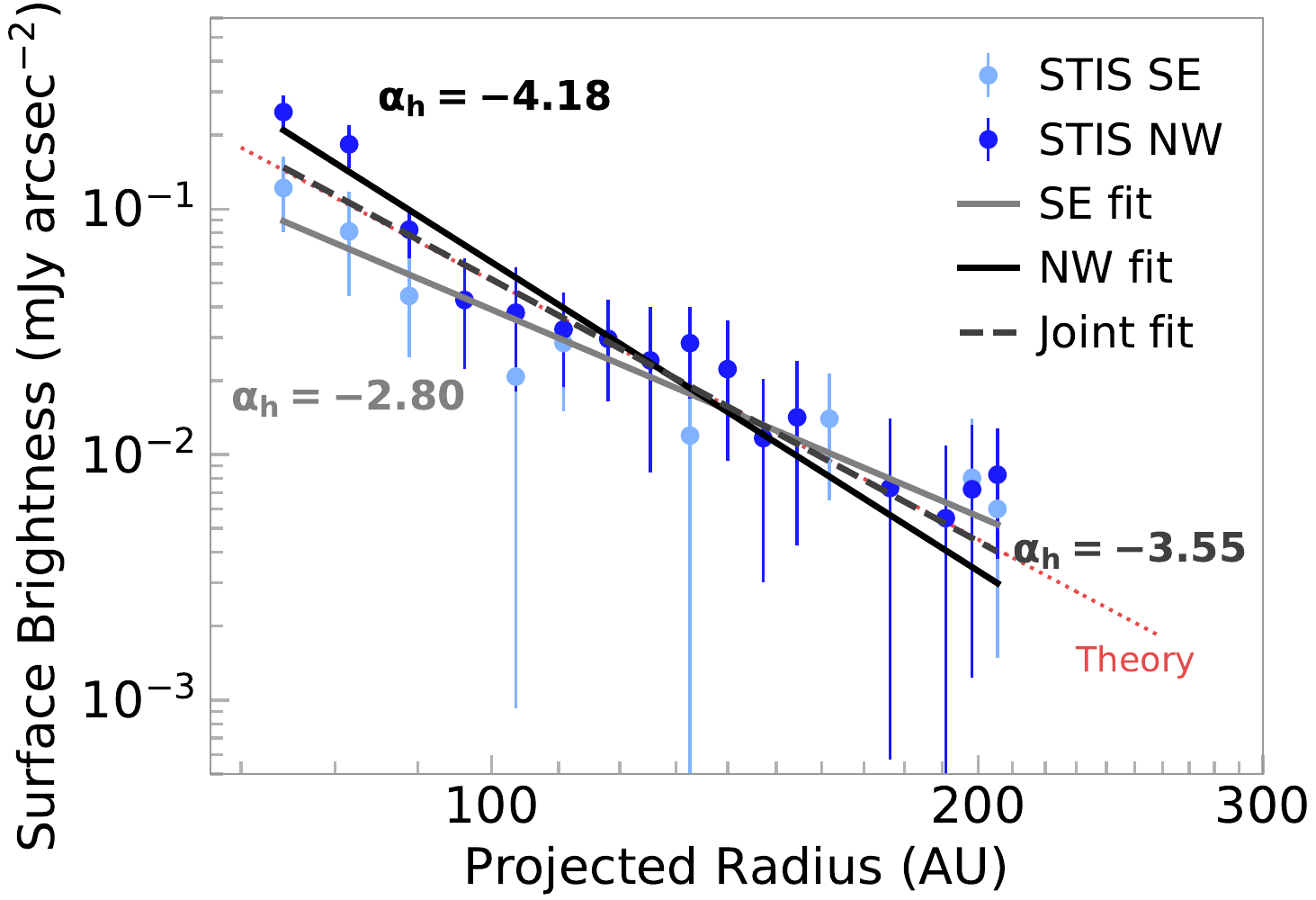}
\caption{Surface brightness radial profiles for the disk's halo measured from the interpolated STIS image. The profiles are divided into the southeast (light shades) and northwest (dark shades) sides of the ring. Errors are 1$\sigma$ uncertainties. The solid lines are the best-fit power law functions for each individual profile and the dashed line is a joint fit to both profiles. All have slopes statistically consistent with the theoretically predicted value of -3.5 (pink dotted line) for collisionally-dominated debris exterior to a planetesimal ``birth ring.'' Surface brightness points that are consistent with zero at the 1$\sigma$ level are neither included in the fit nor plotted.}
\label{fig:halo_sb}
\end{figure}

The halo radial profiles and best-fit power law functions are plotted in Figure \ref{fig:halo_sb}. We found power law indices of $\alpha_h={-}2.80\pm0.36$ in the SE and $\alpha_h={-}4.18\pm0.37$ in the NW. Continuing to assume that the disk is azimuthally symmetric, we also fit a single power law to all measurements from both sides of the halo, and found an index of $\alpha_h={-}3.55\pm0.35$. According to \citet{strubbe2006}, the surface brightnesses of collisionally-dominated debris disks will vary with radius as $\mathrm{SB}\propto r^{-3.5}$ exterior to the dust-producing ``birth ring'' of planetesimals. Indices for individual sides of the halo agree with this predicted value to within $2\sigma$ and the joint index matches it nearly exactly. This is strong evidence that the halo's grains are collisionally produced in the ring and remain gravitationally bound to the star on wide and eccentric orbits. In that case, the brightness in our model from dust on circular orbits with $a \gtrsim 74$ au may simply be a proxy for this eccentric population.

% Halo Radial Profiles.

\subsection{Effects of Mie Scattering on Model Results} \label{sect:disc_mie}

A number of our conclusions about the disk's grain properties are subsequently related to the greater disk environment, which links to planet formation and evolution. However, extracting results for the grain properties required making assumptions about the underlying scattering physics in our disk model, which in our case is based on Mie theory. Therefore, before examining our results more closely we consider how this assumption may have affected them.

Our primary concern is that Mie theory may not accurately reproduce the scattering phase function for grains in debris disks. This would not be surprising given that disk grains, born via collision, are almost certainly not homogeneous perfect spheres like the idealized theory assumes. In particular, \citet{min2016} found that, for equivalent grain radii and porosities, the Mie phase function generally decreases with scattering angle $\theta > 90^\circ$ while a phase function for irregularly shaped aggregate grains is flat or increasing. Therefore, a model exploration that is based on a Mie model may bias parameters controlling grain size, porosity, and composition away from their true values in order to match backward-scattering at large angles seen in comparison data. This would skew the resulting posterior distributions. Aggregate grains, on the other hand, would more naturally produce a backward-scattering component and may prefer different parameter values that are closer to reality. \citet{milli2017} recently pointed out these effects for the debris ring around HR 4796A, for which a Mie model was also incompatible with the scattering phase function. It is important to keep in mind these shortcomings of Mie theory (and other simplified scattering treatments) for the following discussions of grain properties and for future disk modeling efforts.

\subsection{Grain Size and Structure} \label{sect:disc_grain_shape}

Regarding grain structure, our models show a clear preference for a low porosity ($<$12\% at 99.7\% confidence). The disk's polarized intensity is particularly constraining in this regard, as higher porosity tends to increase polarization fraction for a given grain size and composition. Higher porosity also makes grains more forward-scattering, so our view of the ring's back edge in total intensity provides additional constraints. This low porosity is in contrast to models of the AU Mic disk from \citep{graham2007, fitzgerald2007} that require highly porous ($80$\% vacuum) comet-like grains to reproduce its scattering and polarization signatures. This discrepancy may arise from differences in grain size; the \citet{fitzgerald2007} best-fit model with 80\% porosity only contained grains with $0.05\ \micron < a < 3.0\ \micron$ or 3 mm $< a < 6$ mm. Nevertheless, one interpretation of our result favoring compact grains is that little cometary activity occurs in the HD 35841 system. This may be borne out by the non-detection of CO ($J=3-2$) from \citet{moor2011b}. Deeper searches for CO emission and other gas signatures would further investigation on this topic.

In terms of minimum grain size, $a_{min}\approx a_{blow}$ in most of our models. Our uniform Bayesian prior did not contain any information about $a_{blow}$, so agreement of this independent result with the fundamental physics of the system lends credence to the other parameters that we have derived from the light scattering and polarization signatures. It also supports the case for non-porous grains, as higher porosity leads to greater surface area and a larger blowout size. Meanwhile, modeling of some debris disks, like HD 114082 (another F-type star) by \citet{wahhaj2016}, indicate an $a_{min}$ several microns larger than $a_{blow}$. Perhaps this is a true dichotomy resulting from differences in grain porosity and/or shape, or maybe the difference is purely a result of inconsistent modeling methodology. A unified modeling effort applying the same methodology (or better still, a range of methodologies) to multiple disks would provide valuable insight but would also be a substantial undertaking.

Another key aspect of our model grain population is the size distribution slope, which we find to be $2.6\lesssim q\lesssim3.2$ in the 99.7\% confidence interval, with a median value of $q=2.9$. These values lie just below average values of 3.36 \citep{macgregor2016} and 3.15--3.26 \citep{marshall2017} recently estimated for several disks based on their mm-wavelength emission. The thermal brightnesses of those disks are dominated by mm-sized grains, whereas our scattered-light brightness is dominated by micron-sized grains. The fact that our micron-appropriate $q$ values are similar to mm-appropriate $q$ values implies that collisional cascades proceed in a self-similar (read: single power law) way across this size range; that the physics determining particle strengths and particle velocity dispersions does not change qualitatively from millimeter to micron sizes (but see \citealt{strubbe2006} for why the size distribution deviates strongly from a single power law at sizes that are just above the radiation blow-out limit). Future observations of the mm emission from HD 35841 would be useful for verifying that its particle size distribution is indeed characterized by a single power law from millimeters to microns.

\subsection{Grain Composition} \label{sect:disc_composition}

The least constrained of our grain properties are their compositions. Though our models show at least a 2:1 preference for amorphous carbon over astrosilicates and roughly one third of the mass in water ice, the distributions are fairly broad (see Appendix \ref{sect:appA}). Given the resulting uncertainties and influence of Mie approximations, we caution against assigning too much significance to these results.

That said, we can speculate on their implications. For example, the water ice produces a backscattering peak centered around $\theta\approx135^\circ$ that locally enhances the ring's back edge. Future measurements of the phase function with small uncertainties would help confirm this as a real feature. With the low grain porosity implying little cometary activity, the presence of substantial water ice in the disk would need to be explained another way. As for the other materials, the scattering properties of silicate and carbonaceous grains within a single near-IR filter band are very similar apart from albedo. Examples of both cases have been presented in studies of other debris disks. Though difficult, differentiating between a silicate-rich and a carbon-rich disk would be meaningful for the compositions of planets in the system, which presumably formed in and collected material from the same resource pool. A system abundant with carbon and water, two key materials for life on Earth, would be especially interesting from an astrobiology perspective.

% CONCLUSIONS
\section{CONCLUSIONS} \label{sect:conclusions}

With Gemini Planet Imager data we provide the first views of the HD 35841 debris disk that resolve it into a highly inclined dust ring. The ring is detected in the $H$-band in both total intensity and polarized intensity down to projected separations of 12 au. Additional HST STIS broadband optical imaging detects the ring ansae and a smooth dust halo extending outward from the ring.

The ring shows a clear brightness asymmetry along its projected minor axis, which we attribute to the dust grains having a forward-scattering phase function and the ring's west side being the ``front'' side between the star and the observer. We measured the scattering phase function for scattering angles between $22\degr$ and $125\degr$, with upper limits out to $154\degr$. We did the same for the polarized intensity, allowing us to calculate the disk's polarization fraction, which peaks at ${\sim}30$\% near the ring ansae and declines as the scattering angle approaches $0\degr$/$180\degr$.

Coupling the radiative transfer code MCFOST to an MCMC sampler, we compared a large set of scattered-light models with the GPI total intensity and polarized intensity images. This helped us to constrain the ring's inclination to $84.9\degr\substack{+0.2 \\ -0.2}$, inner radius to $60\substack{+1 \\ -2}$ au, and scale height to $2.7\substack{+1.4 \\ -0.3}$ au. It also informed us about the disk's dust properties, indicating a minimum grain size of ${\sim}1.5$ $\micron$ and a size distribution power law index of 2.7--3.0. These models preferred low porosity grains and a total of 0.11--0.19 Earth masses of material in grains sized from 1.5 $\micron$ to 1 mm. They also showed a ${\sim}2$:1 preference for grains to be composed of carbon rather than astrosilicates and be roughly 1/3 water ice by mass.

The scattered-light models, when assessed at visible wavelengths, were also consistent with the STIS image. They formally lacked the outermost part of the broad halo, which we propose is created by radiation pressure pushing grains just larger than $a_{blow}$ onto highly eccentric orbits. Measurements of the radial surface brightness profile of the halo fit this interpretation. Additionally, comparisons of the model's SED with previous measurements suggest that the system contains an inner component that contributes substantial mid-IR flux. We find one possible configuration for this inner component to be a narrow dust ring located at 19--20 au and containing roughly 1/50 the mass of the main ring.

The simplifications involved with our model, such as basing the scattering physics on Mie theory, may have limited our ability to constrain some disk parameters further. This, in turn, limits the statements we can make about the materials present in this circumstellar environment and the dynamical processes at play. Nonetheless, the models presented here provide a self-consistent match to the resolved images and polarimetry in the optical and near-IR and the broadband SED. Promising recent studies have considered approximations of aggregate grain phase functions (\citealt{min2016, tazaki2016}, and references therein) and shown them to agree with a wide collection of observed debris disks \citep{milli2017, hughes2018_review}. Continued development of these models and related codes in sophistication and computational speed will significantly advance our knowledge of grain scattering properties. This will also require commensurate improvements to the fidelity of disk models by fully incorporating physical mechanisms like radiation pressure and grain collisions.

This system remains an interesting target for further observation, as detecting or ruling out the implied inner component would be a potent test for joint imaging+SED modeling predictions. The dust-depleted region inside of the main ring is also a tempting area to search for planets, as we still have few examples of low-mass companions detected at moderate to large separations within resolved debris rings. Such dust-clearing planets represent an important but largely unobserved part of planetary system evolution. Finally, additional multi-wavelength observations that resolve the ring on GPI-like scales would be useful for determining the wavelength dependence of phase functions and polarization fraction, thus providing more points for model-to-data comparison and opportunity to refine our understanding of young circumstellar environments.

\acknowledgments

The authors wish to thank the anonymous referee for helpful suggestions that improved this manuscript. This work is based in part on observations obtained at the Gemini Observatory, which is operated by the Association of Universities for Research in Astronomy, Inc., under a cooperative agreement with the NSF on behalf of the Gemini partnership: the National Science Foundation (United States), the National Research Council (Canada), CONICYT (Chile), Ministerio de Ciencia, Tecnolog\'ia e Innovaci\'on Productiva (Argentina), and Minist\'erio da Ci\^encia, Tecnologia e Inova\c c\~ao (Brazil). Based also in part on observations made with the NASA/ESA Hubble Space Telescope, obtained at the Space Telescope Science Institute, which is operated by the Association of Universities for Research in Astronomy, Inc., under NASA contract NAS 5-26555; these observations are associated with program \#GO-13381. T.M.E., P.K. and J.R.G. thank support from NSF AST-1518332, NASA NNX15AC89G and NNX15AD95G/NEXSS. This work benefited from NASA's Nexus for Exoplanet System Science (NExSS) research coordination network sponsored by NASA's Science Mission Directorate. Portions of this work were also performed under the auspices of the U.S. Department of Energy by Lawrence Livermore National Laboratory under Contract DE-AC52-07NA27344. This work has made use of data from the European Space Agency (ESA) mission {\it Gaia} (\url{https://www.cosmos.esa.int/gaia}), processed by the {\it Gaia} Data Processing and Analysis Consortium (DPAC, \url{https://www.cosmos.esa.int/web/gaia/dpac/consortium}). Funding
for the DPAC has been provided by national institutions, in particular the institutions participating in the {\it Gaia} Multilateral Agreement. This research has made use of the SIMBAD and VizieR databases, operated at CDS, Strasbourg, France.

\software{Gemini Planet Imager Data Pipeline (\citealt{perrin2014_drp, perrin2016_drp}, \url{http://ascl.net/1411.018}), pyKLIP (\citealt{wang2015_pyklip}, \url{http://ascl.net/1506.001}),
emcee (\citealt{foreman-mackey2013}, \url{http://ascl.net/1303.002}), Astropy \citep{astropy2018}, matplotlib \citep{matplotlib2007, matplotlib_v2.0.2}, iPython \citep{ipython2007}, corner (\citealt{corner}, \url{http://ascl.net/1702.002})}.

\facilities{Gemini:South, HST (STIS), Keck:II (NIRC2)}

\appendix

\section{Keck/NIRC2 Reductions}
\label{sect:app_nirc2}

\begin{figure*}[h]
\centering
\includegraphics[width=7.3in]{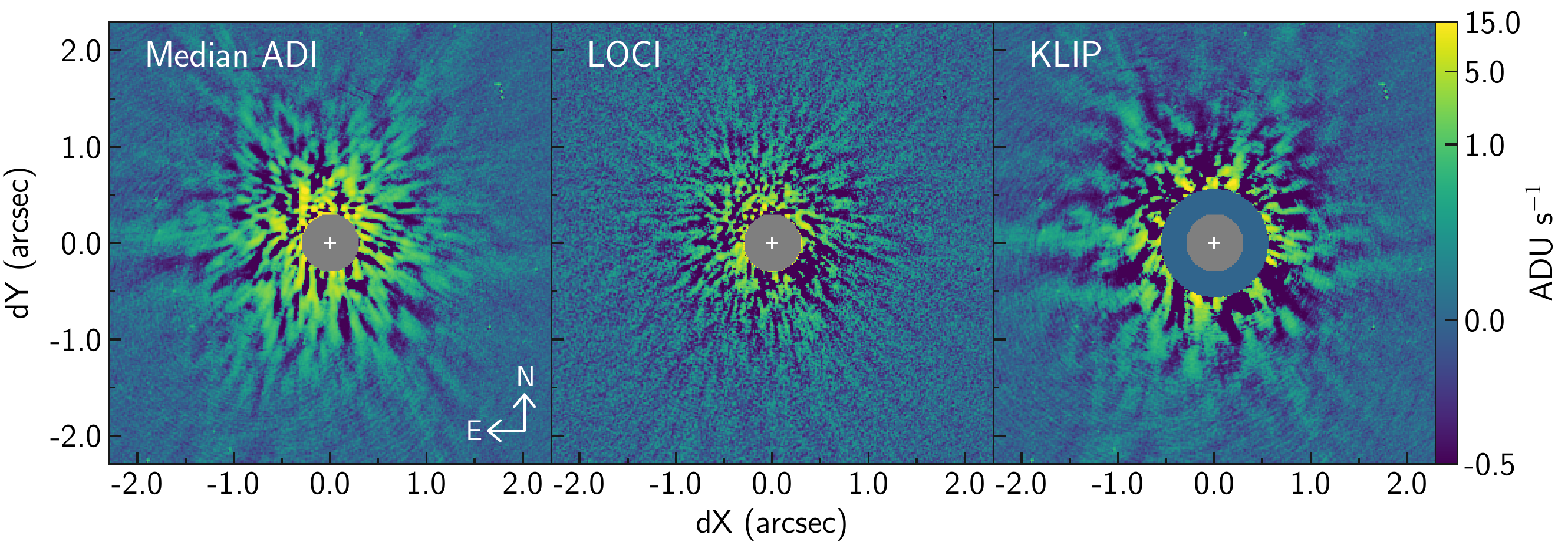}
\caption{NIRC2 $H$-band data reduced with three different PSF-subtraction algorithms: (left) a ``median PSF'' ADI, (middle) LOCI, and (right) \texttt{pyKLIP}. The color is a symmetric logarithmic stretch, the gray circles approximate the size of the focal plane mask ($0\farcs2$ radius), and the white cross marks the star.}
\label{fig:nirc2}
\end{figure*}

The NIRC2 $H$-band data described in Section \ref{sect:nirc2_obs} are shown in Figure \ref{fig:nirc2} reduced with three different PSF-subtraction algorithms. In all cases, we first applied a broad Gaussian highpass filter ($\sigma\sim50$ px) to the images to suppress low frequency background noise. The ``median PSF'' version used a simple median collapse of all images in the data set as the reference PSF, which was then subtracted from all images before the set was averaged across time. The LOCI and \texttt{pyKLIP} reductions used the algorithms described in Section \ref{sect:gpi_obs}. With, LOCI we used 28 annuli in a region of radius = 21--300 px with the number of azimuthal divisions increasing from two at smallest annulus to 14 at the largest, and parameter values of $N_{\delta}=0.5$, $W=4$\,px, $dr=10$\,px, $g=0.1$, and $N_a=250$. For \texttt{pyKLIP}, we used 30 annuli in the 21--300 px region with no azimuthal divisions, a minimum rotation threshold of 10 px, and projection onto the first 25 KL modes.

\clearpage

\section{MCMC POSTERIOR DISTRIBUTIONS} \label{sect:appA}

\begin{figure}[h]
\centering
\includegraphics[width=7.7in, angle=90]{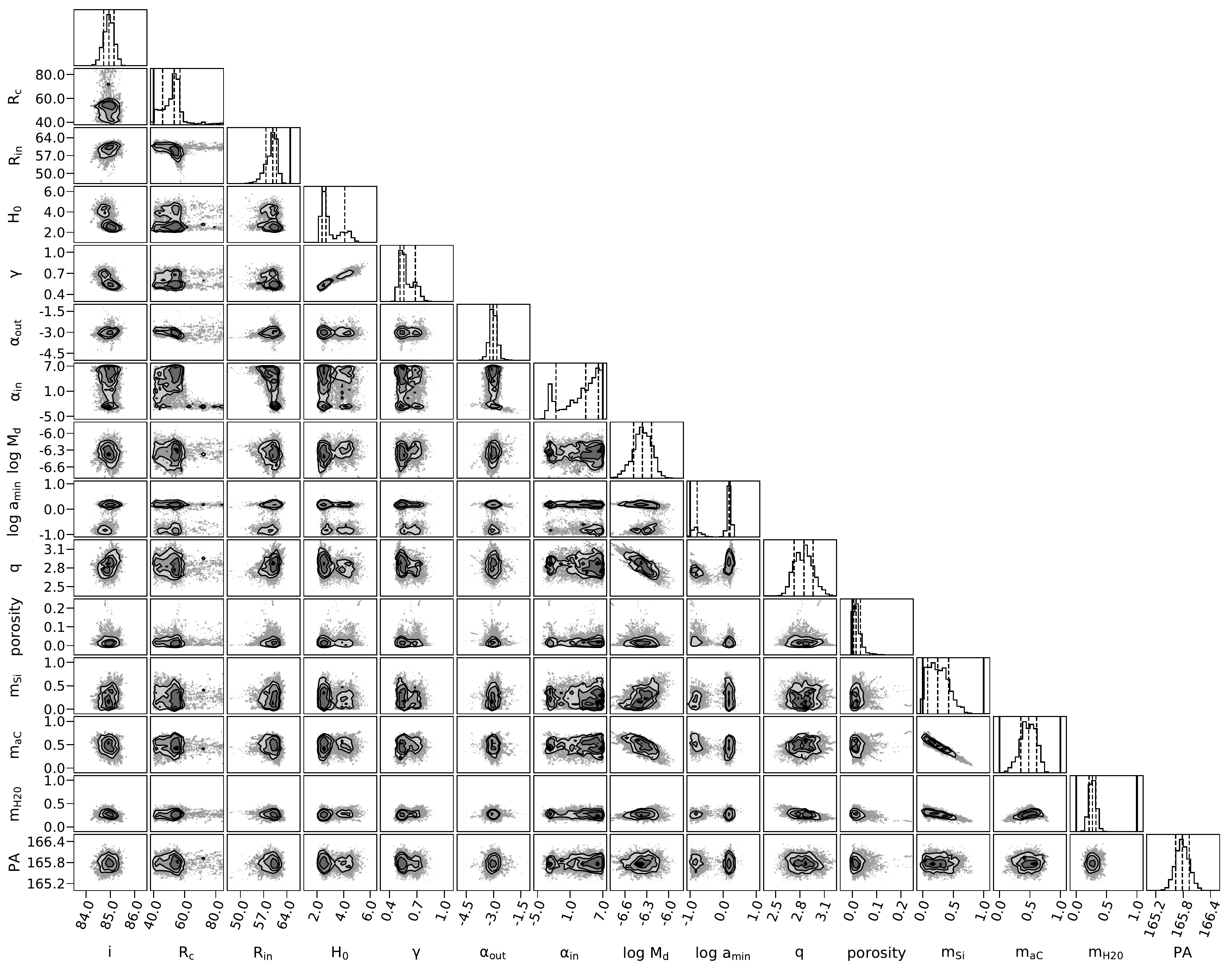}
\caption{Histograms of the marginalized PDF's from the disk model MCMC (Sec \ref{sect:modeling}). Contours in the 2-d histograms mark 39\%, 86\%, and 99\% of the enclosed volume, or the 1$\sigma$, 2$\sigma$, and 3$\sigma$ levels for a 2-d Gaussian density. Dashed lines in the 1-d histograms denote the MCMC 16th, 50th, and 84th percentiles of the marginalized PDF's (left to right) and solid vertical lines denote prior boundaries (these are outside the plotted ranges for some parameters). The units of $i$, $R_{c}$, $R_{in}$, $H_0$, $M_d$, $a_{min}$, and \textit{PA} are deg, au, au, au, $M_\sun$, $\micron$, and deg, respectively. Plot made with the \texttt{corner} Python module \citep{corner}.}
\label{fig:corner}
\end{figure}

\section{HD 35841 SED PHOTOMETRY} \label{sect:appB}

% Observations table.
\begin{table}[h]
\begin{center}
\caption{HD 35841 Photometry}
\label{tab:phot}
\begin{tabular}{l c c c c}
\toprule
Filter & $\lambda_{eff}$ ($\micron$) & Flux (Jy) & Error (Jy) & Ref. \\
\midrule
Johnson B & 0.444 & 0.756 & 0.020 & 1 \\
Johnson V & 0.554 & 0.989 & 0.006 & 1 \\
Sloan $i^\prime$ & 0.763 & 1.10 & 0.06 & 2 \\
Johnson J & 1.25 & 0.979 & 0.009 & 3 \\
Johnson H & 1.63 & 0.760 & 0.021 & 3 \\
Johnson K & 2.19 & 0.482 & 0.022 & 3 \\
WISE W1 & 3.37 & 0.251 & 0.003 & 4 \\
WISE W2 & 4.62 & 0.130 & 0.001 & 4 \\
MIPS 24 & 23.7 & 0.0184 & 0.0007 & 5 \\
MIPS 70 & 71.4 & 0.1721 & 0.0136 & 5 \\
MIPS 160 & 156. & 0.0142 & 0.0142 & 5 \\
SCUBA-2 450 & 450. & 0.035 & 5$\sigma$ up lim & 6 \\
SCUBA-2 850  & 850. & 0.0040 & 5$\sigma$ up lim & 6 \\
\bottomrule
\end{tabular}
\end{center}
References: (1) For B and V, the flux used is the mean of multiple measurements and the error is their standard deviation; \citep{girard2011, nascimbeni2016, mcdonald2017}, (2) \citealt{henden2016}; assumed 5\% error, (3) \citealt{ofek2008}, (4) \citealt{cotten2016}, (5) \citealt{moor2011a}, (6) \citealt{holland2017}.
\end{table}

\bibliographystyle{aasjournal}
\bibliography{disk_exop_refs}

\end{document}